\documentclass[10pt,english]{elsarticle}
\usepackage[T1]{fontenc}
\usepackage[latin9]{inputenc}
\usepackage{geometry}
\usepackage{color}
\usepackage{array}
\usepackage{multirow}
\usepackage{amsmath}
\usepackage{amssymb}
\usepackage{graphicx}

\makeatletter

\providecommand{\tabularnewline}{\\}

\usepackage{scalefnt}
\journal{ }

\makeatother

\usepackage{babel}
\begin{document}

\begin{frontmatter}{}

\title{Quantum decoration transformation for spin models}

\author[{*}]{F. F. Braz, F. C. Rodrigues, S. M. de Souza and Onofre Rojas}

\cortext[a]{email: ors@dfi.ufla.br}

\address{Departamento de Física, Universidade Federal de Lavras, CP 3037,
37200-000, Lavras-MG, Brazil}
\begin{abstract}
It is quite relevant the extension of decoration transformation for
quantum spin models since most of the real materials could be well
described by Heisenberg type models. Here we propose an exact quantum
decoration transformation and also showing interesting properties
such as the persistence of symmetry and the symmetry breaking during
this transformation. Although the proposed transformation, in principle,
cannot be used to map exactly a quantum spin lattice model into another
quantum spin lattice model, since the operators are non-commutative.
However, it is possible the mapping in the \textquotedbl{}classical\textquotedbl{}
limit, establishing an equivalence between both quantum spin lattice
models. To study the validity of this approach for quantum spin lattice
model, we use the Zassenhaus formula, and we verify how the correction
could influence the decoration transformation. But this correction
could be useless to improve the quantum decoration transformation
because it involves the second-nearest-neighbor and further nearest
neighbor couplings, which leads into a cumbersome task to establish
the equivalence between both lattice models. This correction also
gives us valuable information about its contribution, for most of
the Heisenberg type models, this correction could be irrelevant at
least up to the third order term of Zassenhaus formula. This transformation
is applied to a finite size Heisenberg chain, comparing with the exact
numerical results, our result is consistent for weak xy-anisotropy
coupling. We also apply to bond-alternating Ising-Heisenberg chain
model, obtaining an accurate result in the limit of the quasi-Ising
chain.
\end{abstract}

\end{frontmatter}{}

\section{Introduction}

A considerable number of classical decorated Ising models have been
solved using the decoration transformation proposed in the 1950s by
M. E. Fisher\cite{Fisher} and Syozi\cite{syozi}, since that, this
transformation was useful to study decorated Ising lattice in triangular,
honeycomb, Kagomé, and bathroom-tile lattices\cite{houtappel,husimi,syozi-ptp,utiyama},
as well as the Union Jack (centered square)\cite{vaks} and the square
Kagomé\cite{chun-feng} lattice, later pentagonal lattice also was
considered by Urumov\cite{urumov} and by Rojas et al.\cite{mrojas}
among others. Motivated by the above results, later this approach
was generalized in reference \cite{PhyscA-09} for arbitrary spin,
such as the classical-quantum spin models. The decoration transformation
can also be applied to classical-quantum spin models, such as Ising-Heisenberg
models. Several quasi-one-dimensional models have been investigated,
similar to that diamond-like chain\cite{lisni,galisova,strecka-cond-mat09,strecka-conf,strecka06,vadim-10,vadim-nos,vadim-triangl,valvede-diamond}
and references therein, as well as two-dimensional lattice spin models
\cite{maasovska,our-4-6-latt,strecka-2dim-hybrd,galisova-strck,strecka-2dim-quart,strecka-bathroom,strecka-kagome,carlson,pereira-prb78}.
Furthermore, it can be applied even for three-dimensional decorated
systems\cite{3dim-decoration}, this approach can also be applied
combining with Monte Carlo simulation for 3D systems\cite{tsuji,idogaki}. 

Classical decoration transformation could be applied beyond spin models,
such as localized Ising spins and itinerant electrons in two-dimensional
models as discussed by Strecka et al.\cite{strecka-2dim-hybrd}. Later,
the decoration transformation approach has also been applied to spinless
interacting particles, which shows the possibility of application
for interacting electron models\cite{spinless-ferm}. Due to these
meaningful signs of progress, Strecka\cite{strecka-pla} discussed
this transformation in a more detailed fashion, following the approach
proposed in reference \cite{PhyscA-09} for the case of quantum-classical
models. Recently, another interesting transformation\cite{JPA-On-11}
was also suggested to avoid applying several steps of decoration transformations,
by using just one transformation. Alternatively, Derzhko et al.\cite{dersk-strk}
proposed a perturbative approach to study the almost Ising-Heisenberg
diamond chain, by adding a small contribution in XY part.

It is of great importance the extension of classical decoration transformation
for the quantum spin models, because most of the real materials could
be well described by Heisenberg type models. Besides, recent investigations
concerning thermal entanglement have motivated also this mapping such
as q-bits bonded by Heisenberg coupling with finite number of sites.
Thus, quantum decoration transformation could be potentially applied
for small quantum systems in \cite{rigolin,Annls-Hou,annls-Zhang,sha-sha}
and references therein. 

In this paper, we present a pure quantum decoration transformation
for a quantum mixed or decorated quantum spin model into an effective
quantum spin model. The main difference between the classical and
quantum transformation is the non-commutative property; consequently,
the Boltzmann factor becomes an operator. A basic idea of quantum
decoration transformation already has been discussed for a particular
case of diluted Heisenberg model \cite{dunn}. To introduce a quantum
version of decoration transformation for Heisenberg spin models into
an uniform spin-1/2 Heisenberg model, we will follow the basic idea
used by Dunn and Essam\cite{dunn}, as well as by M. E. Fisher \cite{Fisher}
and Syozi\cite{syozi}. 

This paper is organized as follows, in sec. 2 we present the two-leg
quantum decoration transformation, where is included a couple of applications.
In sec. 3 we show the star-triangle decoration transformation, we
also give a couple of applications for the star-triangle transformation.
Whereas, in sec. 4 we discuss how to apply for a quantum spin lattice
model, the correction of the transformation can be obtained using
the Zassenhaus formula. Besides, we apply for finite size Heisenberg
model as well as for bond alternating Ising-Heisenberg model. Finally,
in sec. 5 we give our conclusion and perspectives.

\section{Two leg-star quantum decoration transformation}

To proceed with decoration transformation, we need to extend the Boltzmann
factor\cite{Fisher,syozi,PhyscA-09} to some operator, which can bring
all information about the quantum decorated Hamiltonian.

\begin{figure}
\centering{}\includegraphics[scale=0.6]{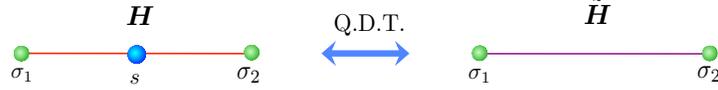}\caption{\label{fig-dec2-leg}Schematic representation of quantum decoration
transformation. Where $s$ corresponds to decorated spin, while $\sigma_{1}$
and $\sigma_{2}$ correspond to Heisenberg spins, $H$ ($\tilde{H}$)
corresponds to decorated (effective) Hamiltonian respectively.}
\end{figure}

Therefore, let us consider a decorated system illustrated in fig.\ref{fig-dec2-leg},
composed of arbitrary quantum operators $\boldsymbol{\sigma}_{1}$,
$\boldsymbol{\sigma}_{2}$, and $\boldsymbol{s}$ is any other quantum
operator or operators called ``decorated operator''. Defining the
operator $\boldsymbol{W}$ as $\boldsymbol{W}={\rm e}^{-\beta\boldsymbol{H}},$
where $\boldsymbol{H}$ is the Hamiltonian of decorated system illustrated
on left side of fig.\ref{fig-dec2-leg}, and $\beta=\frac{1}{k_{B}T}$,
with $k_{B}$ being the Boltzmann constant and $T$ is the absolute
temperature.

Assuming the Hamiltonian $\boldsymbol{H}$'s eigenvalues as $\varepsilon_{n}$,
and denoting the corresponding orthonormal eigenvectors by $|\eta_{n}\rangle$.
Thus, the operator $\boldsymbol{W}$ can be expressed by 
\begin{equation}
\boldsymbol{W}={\rm e}^{-\beta\boldsymbol{H}}=\sum_{n=1}^{\text{dim}(H)}{\rm e}^{-\beta\varepsilon_{n}}|\eta_{n}\rangle\langle\eta_{n}|,
\end{equation}
where $\text{dim}(H)$ means the dimension of the Hamiltonian $\boldsymbol{H}$. 

Multiplying both sides of operator $\boldsymbol{W}$ by the identity
operator $\boldsymbol{1}=\underset{s,\sigma_{1},\sigma_{2}}{\sum}|{}_{s,\sigma_{1},\sigma_{2}}\rangle\langle{}_{s,\sigma_{1},\sigma_{2}}|$,
we have 
\begin{alignat}{1}
\boldsymbol{W}= & \sum_{s,\sigma_{1},\sigma_{2}}\sum_{s',\sigma'_{1},\sigma'_{2}}|{}_{s,\sigma_{1},\sigma_{2}}\rangle\langle{}_{s,\sigma_{1},\sigma_{2}}|{\rm e}^{-\beta\boldsymbol{H}}|{}_{s',\sigma'_{1},\sigma'_{2}}\rangle\langle{}_{s',\sigma'_{1},\sigma'_{2}}|,\\
= & \sum_{s,\sigma_{1},\sigma_{2}}\sum_{s',\sigma'_{1},\sigma'_{2}}\langle{}_{s,\sigma_{1},\sigma_{2}}|{\rm e}^{-\beta\boldsymbol{H}}|{}_{s',\sigma'_{1},\sigma'_{2}}\rangle|{}_{s,\sigma_{1},\sigma_{2}}\rangle\langle{}_{s',\sigma'_{1},\sigma'_{2}}|,\nonumber \\
= & \sum_{n=1}^{\text{dim}(H)}\sum_{s,\sigma_{1},\sigma_{2}}\sum_{s',\sigma'_{1},\sigma'_{2}}{\rm e}^{-\beta\varepsilon_{n}}\langle{}_{s,\sigma_{1},\sigma_{2}}|\eta_{n}\rangle\langle\eta_{n}|{}_{s',\sigma'_{1},\sigma'_{2}}\rangle|{}_{s,\sigma_{1},\sigma_{2}}\rangle\langle{}_{s',\sigma'_{1},\sigma'_{2}}|,\nonumber \\
= & \sum_{n=1}^{\text{dim}(H)}\sum_{s,\sigma_{1},\sigma_{2}}\sum_{s',\sigma'_{1},\sigma'_{2}}{\rm e}^{-\beta\varepsilon_{n}}c_{s,\sigma_{1},\sigma_{2}}^{n*}c_{s',\sigma'_{1},\sigma'_{2}}^{n}|{}_{s,\sigma_{1},\sigma_{2}}\rangle\langle{}_{s',\sigma'_{1},\sigma'_{2}}|,
\end{alignat}
where the coefficients are denoted by $c_{s,\sigma{}_{1},\sigma{}_{2}}^{n}=\langle\eta_{n}|{}_{s,\sigma{}_{1},\sigma{}_{2}}\rangle$
and $c_{s,\sigma_{1},\sigma_{2}}^{n*}=\langle{}_{s,\sigma_{1},\sigma_{2}}|\eta_{n}\rangle$. 

Now, let us calculate the partial trace over decorated quantum operator
$\boldsymbol{s}$, this result is still an operator, called the reduced
operator $\boldsymbol{W}_{r}$, which is expressed as 
\begin{equation}
\boldsymbol{W}_{r}={\rm tr}_{s}\left({\rm e}^{-\beta\boldsymbol{H}}\right)=\sum_{s}\langle_{s}|\boldsymbol{W}|{}_{s}\rangle.\label{eq:par-trace}
\end{equation}
Explicitly, the partial trace in decorated quantum operator becomes
\begin{alignat}{1}
\boldsymbol{W}_{r}= & \sum_{n=1}^{\text{dim}(H)}\sum_{s,\sigma_{1},\sigma_{2}}\sum_{\sigma'_{1},\sigma'_{2}}{\rm e}^{-\beta\varepsilon_{n}}c_{s,\sigma_{1},\sigma_{2}}^{n*}c_{s,\sigma'_{1},\sigma'_{2}}^{n}|{}_{\sigma_{1},\sigma_{2}}\rangle\langle{}_{\sigma'_{1},\sigma'_{2}}|,
\end{alignat}
where we assume the following partial scalar product in $\boldsymbol{s}$:
$|_{\sigma_{1},\sigma_{2}}\rangle=\langle_{s}|{}_{s,\sigma_{1},\sigma_{2}}\rangle$
and $\langle{}_{\sigma_{1},\sigma_{2}}|=\langle{}_{s,\sigma_{1},\sigma_{2}}|_{s}\rangle$.
Using this result, we can rewrite the elements of reduced operator
$\boldsymbol{W}_{r}$ simply as 
\begin{alignat}{1}
r_{\sigma_{1},\sigma_{2};\sigma'_{1},\sigma'_{2}}= & \sum_{n=1}^{\text{dim}(H)}\sum_{s}{\rm e}^{-\beta\varepsilon_{n}}c_{s,\sigma_{1},\sigma_{2}}^{n*}c_{s,\sigma'_{1},\sigma'_{2}}^{n}.\label{eq:r-element}
\end{alignat}

On the other hand, the transformed system is schematically represented
on the right side of fig.\ref{fig-dec2-leg}. Note that the reduced
operator $\boldsymbol{W}_{r}$ only depends on quantum operators $\boldsymbol{\sigma}_{1}$
and $\boldsymbol{\sigma}_{2}$. Therefore, we define another operator
$\tilde{\boldsymbol{W}}={\rm e}^{-\beta\tilde{\boldsymbol{H}}}$,
with $\tilde{\boldsymbol{H}}$ being the Hamiltonian of the transformed
system or simply called the effective Hamiltonian.

Therefore, multiplying both sides of operator $\tilde{\boldsymbol{W}}$
by the identity operator $\boldsymbol{1}=\underset{\sigma_{1},\sigma_{2}}{\sum}|{}_{\sigma_{1},\sigma_{2}}\rangle\langle{}_{\sigma_{1},\sigma_{2}}|$,
we have
\begin{alignat}{1}
\tilde{\boldsymbol{W}}= & \sum_{\sigma_{1},\sigma_{2}}\sum_{\sigma'_{1},\sigma'_{2}}|{}_{\sigma_{1},\sigma_{2}}\rangle\langle{}_{\sigma_{1},\sigma_{2}}|{\rm e}^{-\beta\tilde{\boldsymbol{H}}}|{}_{\sigma'_{1},\sigma'_{2}}\rangle\langle{}_{\sigma'_{1},\sigma'_{2}}|,\nonumber \\
= & \sum_{n=1}^{\text{dim}(\tilde{H})}\sum_{\sigma_{1},\sigma_{2}}\sum_{\sigma'_{1},\sigma'_{2}}{\rm e}^{-\beta\tilde{\varepsilon}_{n}}\tilde{c}_{\sigma_{1},\sigma_{2}}^{n*}\tilde{c}_{\sigma'_{1},\sigma'_{2}}^{n}|{}_{\sigma_{1},\sigma_{2}}\rangle\langle{}_{\sigma'_{1},\sigma'_{2}}|,
\end{alignat}
where $\tilde{c}_{\sigma{}_{1},\sigma{}_{2}}^{n}=\langle\varsigma_{n}|{}_{\sigma{}_{1},\sigma{}_{2}}\rangle$
and $\tilde{c}_{\sigma_{1},\sigma_{2}}^{n*}=\langle{}_{\sigma_{1},\sigma_{2}}|\varsigma_{n}\rangle$.
Similarly, we are considering $\tilde{\varepsilon}_{n}$ and $|\varsigma_{n}\rangle$
being the eigenvalues and eigenvectors of the effective Hamiltonian
$\tilde{\boldsymbol{H}}$.

Furthermore, the elements of $\tilde{\boldsymbol{W}}$ are given by
\begin{alignat}{1}
\tilde{r}_{\sigma_{1},\sigma_{2};\sigma'_{1},\sigma'_{2}}= & \sum_{n=1}^{\text{dim}(\tilde{H})}{\rm e}^{-\beta\tilde{\varepsilon}_{n}}\tilde{c}_{\sigma_{1},\sigma_{2}}^{n*}\tilde{c}_{\sigma'_{1},\sigma'_{2}}^{n}.
\end{alignat}

Imposing the condition of $\boldsymbol{W}_{r}$ and $\tilde{\boldsymbol{W}}$
must be identical, where both of them depend only of $\boldsymbol{\sigma}_{1}$
and $\boldsymbol{\sigma}_{2}$ operators. Consequently, all elements
of $\boldsymbol{W}_{r}$ and $\tilde{\boldsymbol{W}}$ are related
by 
\begin{equation}
r_{\sigma_{1},\sigma_{2};\sigma'_{1},\sigma'_{2}}=\tilde{r}_{\sigma_{1},\sigma_{2};\sigma'_{1},\sigma'_{2}}.
\end{equation}

Surely, this is a natural generalization of classical decoration transformation,
considered initially in reference \cite{Fisher,syozi,PhyscA-09}.

Notice that, performing the trace over all spins of decorated system,
one must obtain the partition function of the system $Z={\rm tr}(\boldsymbol{W})$,
with ${\rm tr}$ denoting the total trace. The operator $\boldsymbol{W}$
divided by $Z$ will be nothing else than the density operator of
the system $\boldsymbol{\rho}=\frac{\boldsymbol{W}}{Z}$.

The above transformation could be applied for a series of Heisenberg
spin models, in this sense, we consider some particular cases to illustrate
this transformation. 

\subsection{Quantum decoration transformation for spin-$(\frac{1}{2},\frac{1}{2},\frac{1}{2})$
XXZ model}

First, let us consider a simple system composed only of three spins,
as represented schematically by fig.\ref{fig-dec2-leg}, with Heisenberg
coupling, then, the Hamiltonian of decorated spin-$(\frac{1}{2},\frac{1}{2},\frac{1}{2})$
XXZ model can be written as
\begin{alignat}{1}
\boldsymbol{H}= & -J\left[s^{x}\left(\sigma_{1}^{x}+\sigma_{2}^{x}\right)+s^{y}\left(\sigma_{1}^{y}+\sigma_{2}^{y}\right)\right]-\Delta s^{z}\left(\sigma_{1}^{z}+\sigma_{2}^{z}\right)-h\left(\sigma_{1}^{z}+\sigma_{2}^{z}\right)-h_{1}s^{z},\label{eq:Ham-2h}
\end{alignat}
 where $\sigma_{i}^{x}=\frac{1}{2}\left[\begin{smallmatrix}0 & 1\\
1 & 0
\end{smallmatrix}\right]$, $\sigma_{i}^{y}=\frac{1}{2}\left[\begin{smallmatrix}0 & -{\rm i}\\
{\rm i} & 0
\end{smallmatrix}\right]$ and $\sigma_{i}^{z}=\frac{1}{2}\left[\begin{smallmatrix}1 & 0\\
0 & -1
\end{smallmatrix}\right]$ are spin-1/2 operators, with $i=\{1,2\}$, and the decorated operator
$s^{\alpha}$ (with $\alpha=\{x,y,z\}$) are also another spin-1/2
operators, same as defined for $\sigma_{i}^{\alpha}$ operators. Whereas,
$J$ is the exchange interaction parameter in $xy$ axes, and $\Delta$
is the anisotropic exchange interaction in $z$ axis. Here $h=gB$
and $h_{1}=g_{1}B$, with $B$ being the magnetic field acting in
$\sigma_{1}^{z}$, $\sigma_{2}^{z}$ and $s^{z}$, while $g_{1}$
($g$) is the Landé g-factor for $s^{z}$ ($\sigma_{1}^{z}$ and $\sigma_{2}^{z}$
), respectively. 

To diagonalize the Hamiltonian $\boldsymbol{H}$, we express the Hamiltonian
in the natural basis $\{|{}_{s,\sigma_{1},\sigma_{2}}\rangle\}=\{|_{\uparrow,+,+}\rangle$,$|_{\uparrow,+,-}\rangle$,
$|_{\uparrow,-,+}\rangle$, $|_{\downarrow,+,+}\rangle$, $|_{\uparrow,-,-}\rangle$,
$|_{\downarrow,+,-}\rangle$, $|_{\downarrow,-,+}\rangle$, $|_{\downarrow,-,-}\rangle\}$,
for details see Appendix \eqref{eq:Hhhh}. 

Hereafter the diagonalization of the Hamiltonian \eqref{eq:Ham-2h}
is performed, the eigenvalues of the Hamiltonian $\boldsymbol{H}$
are given by
\begin{alignat}{4}
\varepsilon_{1}= & -\tfrac{\Delta}{2}-\tfrac{h_{1}}{2}-h,\quad & \varepsilon_{2}= & -\tfrac{h_{1}}{2},\hspace{1.3cm} & \varepsilon_{3}= & \tfrac{\Delta}{4}-\tfrac{h}{2}+\tfrac{\theta}{4},\quad & \varepsilon_{4}= & \tfrac{\Delta}{4}-\tfrac{h}{2}-\tfrac{\theta}{4},\nonumber \\
\varepsilon_{5}= & \tfrac{h_{1}}{2}, & \varepsilon_{6}= & \tfrac{\Delta}{4}+\tfrac{h}{2}+\tfrac{\vartheta}{4}, & \varepsilon_{7}= & \tfrac{\Delta}{4}+\tfrac{h}{2}-\tfrac{\vartheta}{4}, & \varepsilon_{8}= & -\tfrac{\Delta}{2}+\tfrac{h_{1}}{2}+h,
\end{alignat}
with $\theta=\sqrt{\left(\Delta+2h-2h_{1}\right)^{2}+8J^{2}}$ and
$\vartheta=\sqrt{\left(\Delta-2h+2h_{1}\right)^{2}+8J^{2}}$.

Whereas, the corresponding eigenvectors are given respectively by
\begin{alignat}{1}
|\eta_{1}\rangle= & |_{\uparrow,+,+}\rangle,\nonumber \\
|\eta_{2}\rangle= & \tfrac{1}{\sqrt{2}}\left(|_{\uparrow,+,-}\rangle-|{}_{\uparrow,-,+}\rangle\right),\nonumber \\
|\eta_{3}\rangle= & \tfrac{1}{\sqrt{2}}\sin(\phi)|{}_{\uparrow,+,-}\rangle+\tfrac{1}{\sqrt{2}}\sin(\phi)|{}_{\uparrow,-,+}\rangle+\cos(\phi)|{}_{\downarrow,+,+}\rangle,\nonumber \\
|\eta_{4}\rangle= & \tfrac{1}{\sqrt{2}}\cos(\phi)|{}_{\uparrow,+,-}\rangle+\tfrac{1}{\sqrt{2}}\cos(\phi)|{}_{\uparrow,-,+}\rangle-\sin(\phi)|{}_{\downarrow,+,+}\rangle,\nonumber \\
|\eta_{5}\rangle= & \tfrac{1}{\sqrt{2}}\left(|_{\downarrow,+,-}\rangle-|{}_{\downarrow,-,+}\rangle\right),\nonumber \\
|\eta_{6}\rangle= & \tfrac{1}{\sqrt{2}}\sin(\varphi)|{}_{\downarrow,+,-}\rangle+\tfrac{1}{\sqrt{2}}\sin(\varphi)|{}_{\downarrow,-,+}\rangle-\cos(\varphi)|{}_{\uparrow,-,-}\rangle,\nonumber \\
|\eta_{7}\rangle= & \tfrac{1}{\sqrt{2}}\cos(\varphi)|{}_{\downarrow,+,-}\rangle+\tfrac{1}{\sqrt{2}}\cos(\varphi)|{}_{\downarrow,-,+}\rangle+\sin(\varphi)|{}_{\uparrow,-,-}\rangle,\\
|\eta_{8}\rangle= & |_{\downarrow,-,-}\rangle,\nonumber 
\end{alignat}
with 
\begin{alignat}{1}
\phi= & \tan^{-1}\left(\tfrac{\theta-\left(\Delta+2h-2h_{1}\right)}{2\sqrt{2}J}\right),\quad\text{and}\quad-\tfrac{\pi}{2}\leqslant\phi\leqslant\tfrac{\pi}{2},\\
\varphi= & \tan^{-1}\left(\tfrac{\vartheta-\left(\Delta-2h+2h_{1}\right)}{2\sqrt{2}J}\right),\quad\text{and}\quad-\tfrac{\pi}{2}\leqslant\varphi\leqslant\tfrac{\pi}{2}.
\end{alignat}

Furthermore, performing the partial trace in $s$ using eq.\eqref{eq:par-trace},
we obtain the reduced operator $\boldsymbol{W}_{r}$ in terms of natural
basis $\{|{}_{\sigma_{1},\sigma_{2}}\rangle\}=\{|_{+,+}\rangle,|_{+,-}\rangle,|_{-,+}\rangle,|_{-,-}\rangle\}$.
Thus, the operator becomes
\begin{equation}
\boldsymbol{W}_{r}=\left[\begin{array}{cccc}
w_{1} & 0 & 0 & 0\\
0 & \frac{w_{2}+w_{3}}{2} & \frac{w_{2}-w_{3}}{2} & 0\\
0 & \frac{w_{2}-w_{3}}{2} & \frac{w_{2}+w_{3}}{2} & 0\\
0 & 0 & 0 & w_{4}
\end{array}\right],\label{eq:R_r}
\end{equation}
where the elements of $\boldsymbol{W}_{r}$ are obtained from eq.\eqref{eq:r-element},
and result in 
\begin{alignat}{1}
w_{1}=r_{1,1}= & \sum_{n=1}^{\text{dim}(\tilde{H})}\sum_{s=\uparrow,\downarrow}{\rm e}^{-\beta\varepsilon_{n}}|c_{s,+,+}^{n}|^{2},\\
\frac{w_{2}+w_{3}}{2}=r_{2,2}= & \sum_{n=1}^{\text{dim}(\tilde{H})}\sum_{s=\uparrow,\downarrow}{\rm e}^{-\beta\varepsilon_{n}}|c_{s,+,-}^{n}|^{2},\\
\frac{w_{2}-w_{3}}{2}=r_{2,3}= & \sum_{n=1}^{\text{dim}(\tilde{H})}\sum_{s=\uparrow,\downarrow}{\rm e}^{-\beta\varepsilon_{n}}c_{s,+,-}^{n*}c_{s,-,+}^{n},\\
w_{4}=r_{4,4}= & \sum_{n=1}^{\text{dim}(\tilde{H})}\sum_{s=\uparrow,\downarrow}{\rm e}^{-\beta\varepsilon_{n}}|c_{s,-,-}^{n}|^{2}.
\end{alignat}
 After using some algebraic manipulation, we express the elements
of $\boldsymbol{W}_{r}$ in natural basis, as a function of $w_{1},\cdots,w_{4}$,
which are given explicitly as 
\begin{alignat}{1}
w_{1}= & {\rm e}^{-\frac{\beta\left(\Delta-2h+\theta\right)}{4}}\cos^{2}\left(\phi\right)+{\rm e}^{-\frac{\beta\left(\Delta-2h-\theta\right)}{4}}\sin^{2}\left(\phi\right)+{\rm e}^{\frac{\beta\left(\Delta+h_{1}+2h\right)}{2}},\label{eq:w++}\\
w_{2}= & \left({\rm e}^{\frac{\beta{\it \theta}}{4}}\cos^{2}\left(\phi\right)+{\rm e}^{-\frac{\beta{\it \theta}}{4}}\sin^{2}\left(\phi\right)\right){\rm e}^{-\frac{\beta\left(\Delta-2h\right)}{4}}+\left({\rm e}^{-\frac{\beta\vartheta}{4}}\sin^{2}\left(\varphi\right)+{\rm e}^{\frac{\beta{\it \vartheta}}{4}}\cos^{2}\left(\phi\right)\right){\rm e}^{-\frac{\beta\left(\Delta+2h\right)}{4}},\\
w_{3}= & 2\cosh(\frac{\beta h_{1}}{2}),\\
w_{4}= & {\rm e}^{-\frac{\beta\left(\Delta+2h+\vartheta\right)}{4}}\cos^{2}\left(\varphi\right)+{\rm e}^{-\frac{\beta\left(\Delta+2h-\vartheta\right)}{4}}\sin^{2}\left(\varphi\right)+{\rm e}^{\frac{\beta\left(\Delta-h_{1}-2h\right)}{2}}.\label{eq:w--}
\end{alignat}

Alternatively, one can express $\boldsymbol{W}_{r}$ as a function
of orthogonal projection operators (eigenvectors basis) 
\begin{equation}
\boldsymbol{W}_{r}=\sum_{k=1}^{4}w_{k}|\varsigma_{k}\rangle\langle\varsigma_{k}|,\label{eq:W-proj-r}
\end{equation}
 where 
\begin{equation}
|\varsigma_{1}\rangle=|_{++}\rangle,\qquad|\varsigma_{2}\rangle=\tfrac{1}{\sqrt{2}}\left(|_{+-}\rangle+|_{-+}\rangle\right),\qquad|\varsigma_{3}\rangle=\tfrac{1}{\sqrt{2}}\left(|_{+-}\rangle-|_{-+}\rangle\right)\qquad\text{and}\qquad|\varsigma_{4}\rangle=|_{++}\rangle.\label{eq:zetas}
\end{equation}

A similar notation was also used in reference \cite{dunn}, for diluted
Heisenberg model.

On the other hand, let us consider the effective Hamiltonian of spin-$(\frac{1}{2},\frac{1}{2})$
XXZ model as
\begin{alignat}{1}
\tilde{\boldsymbol{H}}= & -\tilde{J}_{0}-\tilde{J}\left(\sigma_{1}^{x}\sigma_{2}^{x}+\sigma_{1}^{y}\sigma_{2}^{y}\right)-\tilde{\Delta}\sigma_{1}^{z}\sigma_{2}^{z}-\tilde{h}\left(\sigma_{1}^{z}+\sigma_{2}^{z}\right),
\end{alignat}
where $\tilde{J}_{0}$ is a ``constant'' energy, $\tilde{J}$ is
the effective exchange parameter between $\boldsymbol{\sigma}_{1}$
and $\boldsymbol{\sigma}_{2}$ in $xy$ axes, $\tilde{\Delta}$ represents
the effective exchange parameter in $z$ axis. Whereas $\tilde{h}=g\tilde{B}$,
$\tilde{B}$ represents the effective external magnetic field in both
operators $\sigma_{1}^{z}$ and $\sigma_{2}^{z}$, with $g$ being
the Landé g-factor.

Writing in the natural basis $\{|{}_{\sigma_{1},\sigma_{2}}\rangle\}$,
the Hamiltonian $\tilde{H}$ becomes
\begin{equation}
\tilde{\boldsymbol{H}}=\left[\begin{array}{cccc}
-\tilde{J}_{0}-\frac{\tilde{\Delta}}{4}-\tilde{h} & 0 & 0 & 0\\
0 & -\tilde{J}_{0}+\frac{\tilde{\Delta}}{4} & -\frac{\tilde{J}}{2} & 0\\
0 & -\frac{\tilde{J}}{2} & -\tilde{J}_{0}+\frac{\tilde{\Delta}}{4} & 0\\
0 & 0 & 0 & -\tilde{J}_{0}-\frac{\tilde{\Delta}}{4}+\tilde{h}
\end{array}\right].\label{eq:Hg12}
\end{equation}

Using the Hamiltonian \eqref{eq:Hg12}, we can define the operator
$\tilde{\boldsymbol{W}}={\rm e}^{-\beta\tilde{\boldsymbol{H}}}$,
in a standard basis $\{|{}_{\sigma_{1},\sigma_{2}}\rangle\}$. Hence,
$\tilde{\boldsymbol{W}}$ matrix is given explicitly by 
\begin{equation}
\tilde{\boldsymbol{W}}=\left[\begin{array}{cccc}
\tilde{w}_{1} & 0 & 0 & 0\\
0 & \frac{\tilde{w}_{2}+\tilde{w}_{3}}{2} & \frac{\tilde{w}_{2}-\tilde{w}_{3}}{2} & 0\\
0 & \frac{\tilde{w}_{2}-\tilde{w}_{3}}{2} & \frac{\tilde{w}_{2}+\tilde{w}_{3}}{2} & 0\\
0 & 0 & 0 & \tilde{w}_{4}
\end{array}\right],\label{eq:Rtil}
\end{equation}
where the elements of $\tilde{\boldsymbol{W}}$ can be expressed regarding
the effective Hamiltonian parameters
\begin{alignat}{5}
\tilde{w}_{1}= & {\rm e}^{-\beta(-\tilde{J}_{0}-\frac{\tilde{\Delta}}{4}-\tilde{h})},\qquad & \tilde{w}_{2}= & {\rm e}^{-\beta(-\tilde{J}_{0}+\frac{\tilde{\Delta}}{4}-\frac{\tilde{J}}{2})},\qquad & \tilde{w}_{3}= & {\rm e}^{-\beta(-\tilde{J}_{0}+\frac{\tilde{\Delta}}{4}+\frac{\tilde{J}}{2})}\quad & \text{and}\quad & \tilde{w}_{4}= & {\rm e}^{-\beta(-\tilde{J}_{0}-\frac{\tilde{\Delta}}{4}+\tilde{h})}.
\end{alignat}

Equivalently, using the projection operator, analogous to operator
eq.\eqref{eq:W-proj-r}, we also have

\begin{equation}
\tilde{\boldsymbol{W}}=\sum_{k=1}^{4}\tilde{w}_{k}|\varsigma_{k}\rangle\langle\varsigma_{k}|,
\end{equation}
where $|\varsigma_{k}\rangle$ are the eigenvectors of $\tilde{\boldsymbol{W}}$,
the same obtained in \eqref{eq:zetas}, and obviously satisfy $[\boldsymbol{W}_{r},\tilde{\boldsymbol{W}}]=0$.

Now, we can impose that the corresponding element must be identical
for both operators and assuming $\tilde{w}_{1}=w_{1}$, $\tilde{w}_{2}=w_{2}$,
$\tilde{w}_{3}=w_{3}$ and $\tilde{w}_{4}=w_{4}$. Besides, this condition
establishes a system of algebraic equation, with 4 unknown parameters
$\tilde{J_{0}}$, $\tilde{J}$, $\tilde{\Delta}$ and $\tilde{h}$,
which can be obtained as a function of decorated Hamiltonian parameters
$J$, $\Delta$, $h$ and $h_{1}$. For instance, let us write just
as a function of $w_{1}$, $w_{2}$, $w_{3}$ and $w_{4}$, which
are explicitly determined by eqs.(\ref{eq:w++}-\ref{eq:w--}). Thus,
solving the algebraic system, we have 

\begin{alignat}{4}
\tilde{J}_{0}= & \frac{1}{4\beta}\ln\left(w_{1}w_{2}w_{3}w_{4}\right),\qquad & \tilde{\Delta}= & \frac{1}{\beta}\ln\left(\tfrac{w_{1}w_{4}}{w_{2}w_{3}}\right),\qquad & \tilde{h}= & \frac{1}{2\beta}\ln\left(\tfrac{w_{1}}{w_{4}}\right)\quad\text{and}\quad & \tilde{J}= & \frac{1}{\beta}\ln\left(\tfrac{w_{2}}{w_{3}}\right).\label{eq:tilde-Js}
\end{alignat}

Finally, one can see the results for $\tilde{J}_{0}$, $\tilde{\Delta}$
and $\tilde{h}$ in terms of $w_{1}$, $w_{2}$, $w_{3}$ and $w_{4}$
are quite similar to the classical decoration transformation\cite{Fisher,syozi,PhyscA-09}.
Whereas the quantum parameter $\tilde{J}$ vanishes when $w_{2}=w_{3}$
following the eq.\eqref{eq:tilde-Js}, because $J=0$ in the original
Hamiltonian.

\subsection{Quantum decoration transformation for spin-$(1,\frac{1}{2},\frac{1}{2})$
XXZ model}

In what follows, we consider a system composed by three spins, one
quantum decorated spin-1 and two quantum spin-1/2. Thus, the Hamiltonian
of decorated spin-$(1,\frac{1}{2},\frac{1}{2})$ XXZ model, is given
by
\begin{alignat}{1}
\boldsymbol{H}= & -J\left[s^{x}\left(\sigma_{1}^{x}+\sigma_{2}^{x}\right)+s^{y}\left(\sigma_{1}^{y}+\sigma_{2}^{y}\right)\right]-\Delta s^{z}\left(\sigma_{1}^{z}+\sigma_{2}^{z}\right).\label{eq:Hmt-ihh}
\end{alignat}

The definition of the model is quite similar to the previous one;
the only difference is $s^{\alpha}$ now becomes $\ensuremath{s^{x}=\frac{1}{\sqrt{2}}\left[\begin{smallmatrix}0 & 1 & 0\\
1 & 0 & 1\\
0 & 1 & 0
\end{smallmatrix}\right]}$, $\ensuremath{s^{y}=\frac{1}{\sqrt{2}}\left[\begin{smallmatrix}0 & -{\rm i} & 0\\
{\rm i} & 0 & -{\rm i}\\
0 & {\rm i} & 0
\end{smallmatrix}\right]}$, $\ensuremath{s^{z}=\left[\begin{smallmatrix}1 & 0 & 0\\
0 & 0 & 0\\
0 & 0 & -1
\end{smallmatrix}\right]}$. Whereas $\sigma_{1}^{\alpha}$ and $\sigma_{2}^{\alpha}$, (with
$\alpha=\{x,y,z\}$) are spin-1/2 operators as defined in previous
section. The Hamiltonian in natural basis $\{|{}_{s,\sigma_{1},\sigma_{2}}\rangle\}$,
becomes a $12\times12$ matrix. The representation of the matrix $\boldsymbol{H}$
is shown in the Appendix eq.\eqref{eq:Hihh}.

After diagonalizing the Hamiltonian \eqref{eq:Hmt-ihh}, we obtain
the following eigenvalues 
\begin{alignat}{4}
\varepsilon_{1}=\varepsilon_{12}= & -\Delta,\qquad & \varepsilon_{2}=\varepsilon_{9}= & J,\qquad\qquad\qquad & \varepsilon_{3}=\varepsilon_{10}= & -J,\nonumber \\
\varepsilon_{4}=\varepsilon_{8}=\varepsilon_{11}= & 0, & \varepsilon_{5}= & \frac{\Delta}{2}+\frac{\Delta}{2\cos(\phi)}, & \varepsilon_{6}= & \frac{\Delta}{2}-\frac{\Delta}{2\cos(\phi)}, & \qquad\varepsilon_{7}= & \Delta,\label{eq:E-eig-ihh}
\end{alignat}
where $\phi=\tan^{-1}\left(\frac{4J}{\sqrt{2}\Delta}\right)$, with
$-\frac{\pi}{2}\leqslant\phi\leqslant\frac{\pi}{2}$. While the corresponding
eigenvector are given in the Appendix eq.\eqref{eq:eig-vc2.2}.

After performing the partial trace over $\boldsymbol{s}$ in eq.\eqref{eq:par-trace},
we obtain the reduced operator $\boldsymbol{W}_{r}$, which has the
same structure to the the eq.\eqref{eq:R_r}, in terms of natural
basis $\{|{}_{\sigma_{1},\sigma_{2}}\rangle\}$. Whereas, $w_{1}$,
$w_{2}$ and $w_{3}$ can be expressed by
\begin{alignat}{1}
w_{1}= & \tfrac{1}{2}{\rm e}^{-\frac{\beta\Delta}{2}}\cosh\left(\tfrac{\beta\Delta}{2\cos(\phi)}\right)-\tfrac{1}{2}{\rm e}^{-\frac{\beta\Delta}{2}}\sinh\left(\tfrac{\beta\Delta}{2\cos(\phi)}\right)\cos(\phi)+\cosh(\beta J)+{\rm e}^{\beta\Delta}+\tfrac{1}{2}{\rm e}^{-\beta\Delta},\nonumber \\
w_{2}= & {\rm e}^{-\frac{\beta\Delta}{2}}\cosh\left(\tfrac{\beta\Delta}{2\cos(\phi)}\right)+{\rm e}^{-\frac{\beta\Delta}{2}}\sinh\left(\tfrac{\beta\Delta}{2\cos(\phi)}\right)\cos(\phi)+2\cosh(\beta J),\nonumber \\
w_{3}= & 3,
\end{alignat}
where for the null magnetic field, we have the symmetry $w_{4}=w_{1}$.

On the other hand, the operator $\tilde{\boldsymbol{W}}$ is given
by eq.\eqref{eq:Rtil}, and its elements can be written as a function
of 
\begin{alignat}{3}
\tilde{w}_{1}= & {\rm e}^{-\beta(-\tilde{J}_{0}-\frac{\tilde{\Delta}}{4})},\qquad & \tilde{w}_{2}= & {\rm e}^{-\beta(-\tilde{J}_{0}+\frac{\tilde{\Delta}}{4}-\frac{\tilde{J}}{2})},\qquad & \tilde{w}_{3}= & {\rm e}^{-\beta(-\tilde{J}_{0}+\frac{\tilde{\Delta}}{4}+\frac{\tilde{J}}{2})},
\end{alignat}
for the limiting case of the null magnetic field, we have the following
relation $\tilde{w}_{4}=\tilde{w}_{1}$.

Hereafter, we can impose that the elements must be identical for both
operators. Analogous to the previous case, we have three parameters
$\tilde{J_{0}}$, $\tilde{J}$, and $\tilde{\Delta}$ to be determined
and three algebraic equations. Then, we can obtain all unknown parameters
$\tilde{J_{0}}$, $\tilde{J}$ and $\tilde{\Delta}$ in a transformed
system as a function of decorated Hamiltonian parameters $J$ and
$\Delta$. Thus, solving the algebraic system we have
\begin{alignat}{3}
\tilde{J}_{0}= & \frac{1}{4\beta}\ln\left[w_{1}^{2}w_{2}w_{3}\right],\qquad & \tilde{\Delta}= & \frac{1}{\beta}\ln\left(\tfrac{w_{1}^{2}}{w_{2}w_{3}}\right),\qquad & \tilde{J}= & \frac{1}{\beta}\ln\left(\tfrac{w_{2}}{w_{3}}\right).
\end{alignat}

Note that, this result can also be obtained directly from eqs. \eqref{eq:tilde-Js}
assuming $w_{1}=w_{4}$. For the case of $w_{2}=w_{3}$, we have $\tilde{J}=0$,
which corresponds to the classical decoration transformation.

Therefore, to transform the isotropic Heisenberg model into another
effective isotropic Heisenberg model, we need to impose the following
relation 
\begin{equation}
w_{1}w_{4}=w_{2}^{2}.\label{eq:2-ws-rel}
\end{equation}

The Isotropic ($\Delta=J$) Heisenberg model under null magnetic field
can be mapped into another effective isotropic ($\tilde{\Delta}=\tilde{J}$)
Heisenberg model (XXX$\leftrightarrow$XXX), although this symmetry
is breaking when the model is under magnetic field. Thus, the isotropic
($\Delta=J$) Heisenberg model will be mapped into another effective
anisotropic ($\tilde{\Delta}\ne\tilde{J}$) Heisenberg model (XXX$\leftrightarrow$XXZ).
Several variants of Heisenberg type models can be mapped into another
effective Heisenberg type models.

Not all models can be mapped into another model with its original
symmetry. Such as the XY model, this model cannot be mapped into another
effective XY model, but into a more general XYZ model (XY$\leftrightarrow$XYZ),
not discussed here\cite{xy}. 

\section{Star-triangle quantum decoration transformation}

Now let us consider another quite interesting quantum system with
3-leg star, a decorated operator is located in the center of the star
denoted by $\boldsymbol{s}$ operator called \textquotedbl{}decorated
operator\textquotedbl{}, and in each leg is distributed the operators
$\boldsymbol{\sigma}_{1}$, $\boldsymbol{\sigma}_{2}$ and $\boldsymbol{\sigma}_{3}$.

\begin{figure}
\centering{}\includegraphics[scale=0.5]{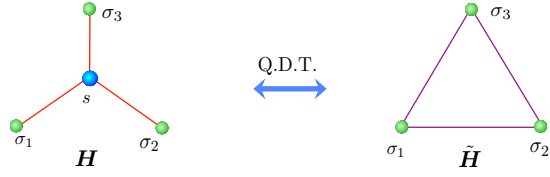}\caption{\label{fig-dec3-leg} Schematic representation of star-triangle quantum
decoration transformation. Where $s$ corresponds to decorated spin,
while $\sigma_{1}$, $\sigma_{2}$ and $\sigma_{3}$ correspond to
Heisenberg spins, $H$ ($\tilde{H}$) corresponds to decorated (effective)
Hamiltonian respectively.}
\end{figure}

Defining the operator $\boldsymbol{W}={\rm e}^{-\beta\boldsymbol{H}},$
where $\boldsymbol{H}$ is the Hamiltonian of decorated system illustrated
on left side of fig.\ref{fig-dec3-leg}. Assuming the Hamiltonian
$\boldsymbol{H}$'s eigenvalues as $\varepsilon_{n}$, and the corresponding
orthonormal eigenvectors by $|\eta_{n}\rangle$, with $n=\{1,...,{\rm dim}(H)\}$. 

Explicitly, the operator $\boldsymbol{W}$ can be expressed by
\begin{equation}
\boldsymbol{W}={\rm e}^{-\beta\boldsymbol{H}}=\sum_{n=1}^{\text{dim}(H)}{\rm e}^{-\beta\varepsilon_{n}}|\eta_{n}\rangle\langle\eta_{n}|.
\end{equation}
Multiplying both sides of operator $\boldsymbol{W}$ by identity operator
$\boldsymbol{1}=\underset{s,\{\sigma_{i}\}}{\sum}|{}_{s,\{\sigma_{i}\}}\rangle\langle{}_{s,\{\sigma_{i}\}}|$,
and using for convenience the following notation $\{\sigma_{i}\}=\{\sigma_{1},\sigma_{2},\sigma_{3}\}$,
the summation runs over $s$, $\sigma_{1}$, $\sigma_{2}$, and $\sigma_{3}$.
Thus, the operator $\boldsymbol{W}$ becomes
\begin{alignat}{1}
\boldsymbol{W}= & \sum_{n=1}^{\text{dim}(H)}\sum_{s,\{\sigma_{i}\}}\sum_{s',\{\sigma'_{i}\}}{\rm e}^{-\beta\varepsilon_{n}}c_{s,\{\sigma_{i}\}}^{n*}c_{s',\{\sigma'_{i}\}}^{n}|{}_{s,\{\sigma_{i}\}}\rangle\langle{}_{s',\{\sigma'_{i}\}}|,
\end{alignat}
where the scalar products are $c_{s,\{\sigma{}_{i}\}}^{n}=\langle\eta_{n}|{}_{s,\{\sigma{}_{i}\}}\rangle$
and $c_{s,\{\sigma{}_{i}\}}^{n*}=\langle{}_{s,\{\sigma{}_{i}\}}|\eta_{n}\rangle$.

Performing the partial trace over $s$, given by eq \eqref{eq:par-trace},
we have the reduced operator
\begin{alignat}{1}
\boldsymbol{W}_{r}= & \sum_{n=1}^{\text{dim}(H)}\sum_{s,\{\sigma_{i}\}}\sum_{\{\sigma'_{i}\}}{\rm e}^{-\beta\varepsilon_{n}}c_{s,\{\sigma_{i}\}}^{n*}c_{s',\{\sigma'_{i}\}}^{n}|{}_{\{\sigma_{i}\}}\rangle\langle{}_{\{\sigma'_{i}\}}|,\label{eq:Wr}
\end{alignat}
with the scalar products given by $|_{\{\sigma{}_{i}\}}\rangle=\langle_{s}|{}_{s,\{\sigma{}_{i}\}}\rangle$
and $\langle{}_{\{\sigma{}_{i}\}}|=\langle{}_{s,\{\sigma{}_{i}\}}|_{s}\rangle$.
Using this representation, we can rewrite the elements of reduced
operator $\boldsymbol{W}_{r}$ conveniently as
\begin{alignat}{1}
r_{\{\sigma{}_{i}\};\{\sigma'_{i}\}}= & \sum_{n=1}^{\text{dim}(H)}\sum_{s}{\rm e}^{-\beta\varepsilon_{n}}c_{s,\{\sigma{}_{i}\}}^{n*}c_{s,\{\sigma'_{i}\}}^{n}.\label{eq:r-elem}
\end{alignat}

On the other hand, the transformed system is represented schematically
on the right side of fig.\ref{fig-dec3-leg}. Analogously, we define
another operator $\tilde{\boldsymbol{W}}$, as $\tilde{\boldsymbol{W}}={\rm e}^{-\beta\tilde{\boldsymbol{H}}}$,
where $\tilde{\boldsymbol{H}}$ is the Hamiltonian of the transformed
system called as effective Hamiltonian.

Multiplying the operator $\tilde{\boldsymbol{W}}$ by the identity
operator $\boldsymbol{1}=\underset{\{\sigma_{i}\}}{\sum}{\rm e}^{-\beta\tilde{\boldsymbol{H}}}|{}_{\{\sigma{}_{i}\}}\rangle\langle{}_{\{\sigma{}_{i}\}}|$,
we have 
\begin{alignat}{1}
\tilde{\boldsymbol{W}}= & \sum_{n=1}^{4}\sum_{\{\sigma{}_{i}\}}\sum_{\{\sigma'_{i}\}}{\rm e}^{-\beta\tilde{\varepsilon}_{n}}\tilde{c}_{\{\sigma{}_{i}\}}^{n*}\tilde{c}_{\{\sigma'_{i}\}}^{n}|{}_{\{\sigma{}_{i}\}}\rangle\langle{}_{\{\sigma'_{i}\}}|,\label{eq:W tilde}
\end{alignat}
with scalar products being $\tilde{c}_{\{\sigma{}_{i}\}}^{n}=\langle\varsigma_{n}|{}_{\{\sigma{}_{i}\}}\rangle$
and $\tilde{c}_{\{\sigma{}_{i}\}}^{n*}=\langle{}_{\{\sigma{}_{i}\}}|\varsigma_{n}\rangle$.
Thus, the elements of $\tilde{\boldsymbol{W}}$ are given by
\begin{alignat}{1}
\tilde{r}_{\{\sigma{}_{i}\};\{\sigma'_{i}\}}= & \sum_{n=1}^{4}{\rm e}^{-\beta\tilde{\varepsilon}_{n}}\tilde{c}_{\{\sigma{}_{i}\}}^{n*}\tilde{c}_{\{\sigma'_{i}\}}^{n}.\label{eq:rt-elem}
\end{alignat}

After the partial trace is performed in $\boldsymbol{s}$, the elements
of $\tilde{\boldsymbol{W}}$ must be identical to $\boldsymbol{W}_{r}$.
Certainly, this is a natural generalization of a classical star-triangle
decoration transformation\cite{Fisher}. This means that, all elements
of $\boldsymbol{W}_{r}$ and $\tilde{\boldsymbol{W}}$ composed only
of $\boldsymbol{\sigma}_{1}$, $\boldsymbol{\sigma}_{2}$ and $\boldsymbol{\sigma}_{3}$
must satisfy the following relation 
\begin{equation}
r_{\{\sigma{}_{i}\};\{\sigma'_{i}\}}=\tilde{r}_{\{\sigma{}_{i}\};\{\sigma'_{i}\}}.
\end{equation}

Surely, this result can be straightforwardly generalized for a $n$-leg
star-polygon transformation, we just need to extend our spin notation
to $\{\sigma_{i}\}=\{\sigma_{1},\sigma_{2},\ldots\sigma_{n}\}$. Although,
this transformation will involve next nearest and further nearest
neighbor coupling terms, similar to that discussed in reference \cite{PhyscA-09}.

In the following subsections, we will give a couple of examples to
show how this transformation works.

\subsection{Quantum decoration transformation for spin-$(\frac{1}{2},\frac{1}{2},\frac{1}{2},\frac{1}{2})$
XXZ model}

Let us transform a typical star shape system, composed of spin-1/2
Heisenberg model. Whose Hamiltonian of star model can be considered
as decorated system with spin-$(\frac{1}{2},\frac{1}{2},\frac{1}{2},\frac{1}{2})$
XXZ model, the first spin corresponds to the central spin, thus we
have 
\begin{alignat}{1}
\boldsymbol{H}= & -J\left[s^{x}\left(\sigma_{1}^{x}+\sigma_{2}^{x}+\sigma_{3}^{x}\right)+s^{y}\left(\sigma_{1}^{y}+\sigma_{2}^{y}+\sigma_{3}^{y}\right)\right]-\Delta s^{z}\left(\sigma_{1}^{z}+\sigma_{2}^{z}+\sigma_{3}^{z}\right),\label{eq:Ham-3h}
\end{alignat}
where $\sigma_{i}^{\alpha}$ and $s^{\alpha}$ (with $\alpha=\{x,y,z\}$
and $i=\{1,2,3\}$) are spin-1/2 operators, as well as $J$ and $\Delta$
are Heisenberg parameters as defined in eq.\eqref{eq:Ham-2h}.

To diagonalize the Hamiltonian \eqref{eq:Ham-3h}, we can express
the Hamiltonian in natural basis $\{|{}_{s,\sigma_{1},\sigma_{2},\sigma_{3}}\rangle\}$
as $16\times16$ matrix. 

Therefore, the Hamiltonian \eqref{eq:Ham-3h} eigenvalues become 
\begin{alignat}{3}
\varepsilon_{1}=\varepsilon_{2}= & -\frac{3\Delta}{4},\qquad & \varepsilon_{3}=\varepsilon_{4}=\varepsilon_{5}=\varepsilon_{6}= & -\frac{\Delta}{4},\qquad & \varepsilon_{7}=\varepsilon_{8}= & \frac{\Delta}{4}\pm J,\nonumber \\
\varepsilon_{9}=\varepsilon_{10}=\varepsilon_{11}=\varepsilon_{12}= & \frac{\Delta}{4}\pm\frac{J}{2}, & \varepsilon_{13}=\varepsilon_{14}=\varepsilon_{15}=\varepsilon_{16}= & \frac{\Delta}{4}\pm\frac{\Delta}{2\cos(\phi)},\label{eq:Eig-E-3h}
\end{alignat}
where $\phi=\tan^{-1}\left(\frac{\sqrt{3}J}{\Delta}\right)$ with
$-\frac{\pi}{2}\leqslant\phi\leqslant\frac{\pi}{2}$.

The corresponding eigenvectors are expressed in Appendix eq.\eqref{eq:eigvct-3h}.
Once we have calculated the partial trace using the eq.\eqref{eq:Wr},
we obtain the reduced operator $\boldsymbol{W}_{r}$ as 
\begin{alignat}{1}
\boldsymbol{W}_{r}= & \left[\begin{array}{cccccccc}
w_{1} & 0 & 0 & 0 & 0 & 0 & 0 & 0\\
0 & \frac{w_{2}+2w_{3}}{3} & \frac{w_{2}-w_{3}}{3} & \frac{w_{2}-w_{3}}{3} & 0 & 0 & 0 & 0\\
0 & \frac{w_{2}-w_{3}}{3} & \frac{w_{2}+2w_{3}}{3} & \frac{w_{2}-w_{3}}{3} & 0 & 0 & 0 & 0\\
0 & \frac{w_{2}-w_{3}}{3} & \frac{w_{2}-w_{3}}{3} & \frac{w_{2}+2w_{3}}{3} & 0 & 0 & 0 & 0\\
0 & 0 & 0 & 0 & \frac{w_{2}+2w_{3}}{3} & \frac{w_{2}-w_{3}}{3} & \frac{w_{2}-w_{3}}{3} & 0\\
0 & 0 & 0 & 0 & \frac{w_{2}-w_{3}}{3} & \frac{w_{2}+2w_{3}}{3} & \frac{w_{2}-w_{3}}{3} & 0\\
0 & 0 & 0 & 0 & \frac{w_{2}-w_{3}}{3} & \frac{w_{2}-w_{3}}{3} & \frac{w_{2}+2w_{3}}{3} & 0\\
0 & 0 & 0 & 0 & 0 & 0 & 0 & w_{1}
\end{array}\right],\label{eq:w3-org}
\end{alignat}
where the elements can be written as a function of
\begin{alignat}{1}
w_{1}= & {\rm e}^{-\frac{\beta\Delta}{4}}\cosh\left(\tfrac{\beta\Delta}{2\cos(\phi)}\right)-{\rm e}^{-\frac{\beta\Delta}{4}}\sinh\left(\tfrac{\beta\Delta}{2\cos(\phi)}\right)\cos(\phi)+{\rm e}^{\frac{3\beta\Delta}{4}},\\
w_{2}= & {\rm e}^{-\frac{\beta\Delta}{4}}\cosh\left(\tfrac{\beta\Delta}{2\cos(\phi)}\right)+{\rm e}^{-\frac{\beta\Delta}{4}}\sinh\left(\tfrac{\beta\Delta}{2\cos(\phi)}\right)\cos(\phi)+{\rm e}^{-\frac{\beta\Delta}{4}}\cosh\left(\beta J\right),\\
w_{3}= & {\rm e}^{-\frac{\beta\Delta}{4}}\cosh\left(\tfrac{\beta J}{2}\right)+{\rm e}^{\tfrac{\beta\Delta}{4}}.
\end{alignat}

On the other hand, the effective spin-1/2 Heisenberg model in a triangle
system can be expressed by
\begin{alignat}{1}
\tilde{\boldsymbol{H}}= & -\tilde{J}_{0}-\tilde{J}\left(\sigma_{1}^{x}\sigma_{2}^{x}+\sigma_{1}^{y}\sigma_{2}^{y}+\sigma_{2}^{x}\sigma_{3}^{x}+\sigma_{2}^{y}\sigma_{3}^{y}+\sigma_{3}^{x}\sigma_{1}^{x}+\sigma_{3}^{y}\sigma_{1}^{y}\right)-\tilde{\Delta}\left(\sigma_{1}^{z}\sigma_{2}^{z}+\sigma_{2}^{z}\sigma_{3}^{z}+\sigma_{3}^{z}\sigma_{1}^{z}\right),\label{eq:tilde-H3}
\end{alignat}
 where $\tilde{J}_{0}$ is a ``constant'' energy, $\tilde{J}$ is
the exchange parameter between $\boldsymbol{\sigma}_{1}$, $\boldsymbol{\sigma}_{2}$
and $\boldsymbol{\sigma}_{3}$ in $xy$ axes, $\tilde{\Delta}$ represents
the exchange parameter in $z$ axis. 

Thus, the elements of matrix $\tilde{\boldsymbol{W}}$ is given by
\eqref{eq:W tilde}, which can be written as a function of, 
\begin{alignat}{3}
\tilde{w}_{1}= & {\rm e}^{\beta(\tilde{J}_{0}+\frac{3\tilde{\Delta}}{4})},\qquad & \tilde{w}_{2}= & {\rm e}^{\beta(\tilde{J}_{0}+\tilde{J}-\frac{\tilde{\Delta}}{4})},\qquad & \tilde{w}_{3}= & {\rm e}^{\beta(\tilde{J}_{0}-\frac{\tilde{J}}{2}-\frac{\tilde{\Delta}}{4})}.\label{eq:TS-tw1}
\end{alignat}

Analogously to the previous case, imposing the condition $\tilde{w}_{1}=w_{1}$,
$\tilde{w}_{2}=w_{2}$, $\tilde{w}_{3}=w_{3}$, we find an algebraic
systems equations. Solving the system equations, we have
\begin{alignat}{3}
\tilde{J}_{0}= & \frac{1}{4\beta}\ln\left[w_{1}w_{2}w_{3}^{2}\right],\qquad & \tilde{\Delta}= & \frac{1}{3\beta}\ln\left(\tfrac{w_{1}^{3}}{w_{2}w_{3}^{2}}\right),\qquad & \tilde{J}= & \frac{2}{3\beta}\ln\left(\tfrac{w_{2}}{w_{3}}\right).\label{eq:TS-tJ0}
\end{alignat}
For the case of $w_{2}=w_{3}$, the eq. \eqref{eq:w3-org} becomes
a diagonal matrix, because $\tilde{J}=0$ or $J=0$ dropping to classical
decoration transformation.

\subsection{Quantum decoration transformation for spin-$(1,\frac{1}{2},\frac{1}{2},\frac{1}{2})$
XXZ model}

Here, we study a typical star structure with central spin-1, and other
spins are spin-1/2 particles. Thus, the system is described by the
Hamiltonian of spin-$(1,\frac{1}{2},\frac{1}{2},\frac{1}{2})$ XXZ
model, which is given by 
\begin{alignat}{1}
\boldsymbol{H}= & -\left[Js^{x}\left(\sigma_{1}^{x}+\sigma_{2}^{x}+\sigma_{3}^{x}\right)+Js^{y}\left(\sigma_{1}^{y}+\sigma_{2}^{y}+\sigma_{3}^{y}\right)+\Delta s^{z}\left(\sigma_{1}^{z}+\sigma_{2}^{z}+\sigma_{3}^{z}\right)\right],\label{eq:TS-1}
\end{alignat}
writing the Hamiltonian \eqref{eq:TS-1} in standard basis $\{|{}_{s,\sigma_{1},\sigma_{2},\sigma_{3}}\rangle\}$,
we have a matrix with dimension $24\times24$, this matrix becomes
large enough to write explicitly with several zero elements.

For the anisotropic case ($\Delta\ne J$), the eigenvalues of the
Hamiltonian \eqref{eq:TS-1} involve cubic algebraic equations, one
can solve this one analytically. However, here we only focus on the
isotropic case ($\Delta=J$), and its solution just involves quadratic
algebraic equations. 

Furthermore, using eq.\eqref{eq:Wr}, we can obtain the reduced operator
$\boldsymbol{W}_{r}$, which has exactly the same structure to the
eq.\eqref{eq:w3-org}. Then, only $w_{1}$, $w_{2}$ and $w_{3}$
are defined by the following expressions 
\begin{alignat}{3}
w_{1}= & \tfrac{1}{2}{\rm e}^{-\frac{\beta5\Delta}{2}}+\tfrac{3}{2}{\rm e}^{\frac{\beta3\Delta}{2}}+{\rm e}^{-\beta\Delta},\qquad & w_{2}= & \tfrac{1}{6}{\rm e}^{-\frac{\beta5\Delta}{2}}+\tfrac{1}{2}{\rm e}^{\frac{\beta3\Delta}{2}}+\tfrac{1}{2}{\rm e}^{-\beta\Delta}+\tfrac{1}{3}{\rm e}^{\frac{\beta\Delta}{2}},\qquad & w_{3}= & \tfrac{1}{2}{\rm e}^{-\beta\Delta}+{\rm e}^{\frac{\beta\Delta}{2}}.\label{eq:TS-w1}
\end{alignat}

On the other side, the Hamiltonian of effective spin-1/2 Heisenberg
model in a triangle system is given by \eqref{eq:tilde-H3}, because
the effective model is exactly the same to the previous application. 

Imposing the condition $\tilde{\boldsymbol{W}}=\boldsymbol{W}_{r}$,
whose matrix structure is given by \eqref{eq:w3-org}. Therefore,
the effective parameters are given by \eqref{eq:TS-tJ0}, and the
explicit expressions of $w_{1}$, $w_{2}$ and $w_{3}$ are given
by eq.\eqref{eq:TS-w1}. 

It is quite remarkable the extension of decoration transformation
for quantum spin models since most of the real materials could be
well described by Heisenberg type models. Furthermore, the quantum
decoration transformation could be exactly applied to small quantum
systems, such as coupled spin systems\cite{rigolin,Annls-Hou,annls-Zhang,sha-sha}
among other models.

In addition, the decoration transformation can be applied, for several
other models. However, we must be careful in applying this approach,
because not all Hamiltonians satisfy its corresponding original Hamiltonian
symmetry. It is worth to mention also, that this transformation cannot
be applied naively for lattice models, due to non-commuting operators. 

\section{Decoration transformation for quantum lattice models}

Obviously, the decoration transformation discussed previously cannot
be applied naively for quantum spin lattice models. Contrary to the
classical spin decoration transformation which can be applied exactly
to lattice spin models. 

\subsection{Quantum decoration transformation correction using Zassenhaus formula}

For quantum decoration transformation, the operators are no longer
commutative operators, because immediately arises a second nearest
neighbor and further nearest neighbors, leading to a very cumbersome
Hamiltonian turning its solution in a tricky task, and the spirit
of the decoration transformation is completely lost.

For instance, without lose its generality, let us consider a quantum
chain model given by 
\begin{equation}
\boldsymbol{\mathcal{H}}=\sum_{i=1}^{2N}\boldsymbol{H}_{i,i+1}=\boldsymbol{H}_{1,2}+\boldsymbol{H}_{2,3}+\boldsymbol{H}_{3,4}\cdots,
\end{equation}
 with open boundary condition and $2N+1$ sites, and assuming all
even sites as decorated spins. 

Therefore, grouping the Hamiltonian as follows 
\begin{alignat}{1}
\boldsymbol{\mathcal{H}} & =\left(\boldsymbol{H}_{1,2}+\boldsymbol{H}_{2,3}\right)+\left(\boldsymbol{H}_{3,4}+\boldsymbol{H}_{4,5}\right)+\left(\boldsymbol{H}_{5,6}+\boldsymbol{H}_{6,7}\right)+\cdots,\\
 & =\boldsymbol{H}_{1,3}+\boldsymbol{H}_{3,5}+\boldsymbol{H}_{5,7}+\ldots,
\end{alignat}
denoting $\boldsymbol{H}_{2i-1,2i+1}=\boldsymbol{H}_{2i-1,2i}+\boldsymbol{H}_{2i,2i+1}$.
In this notation all even sites are considered as decorated spins,
then formally the system Hamiltonian becomes
\begin{equation}
\boldsymbol{\mathcal{H}}=\sum_{i=1}^{N}\boldsymbol{H}_{2i-1,2i+1}.\label{eq:Ham-tot}
\end{equation}

Now we want to apply the decoration transformation for whole lattice
system, then we can use the Zassenhaus formula\cite{casas} (${\rm e}^{\boldsymbol{X}+\boldsymbol{Y}}={\rm e}^{\boldsymbol{X}}{\rm e}^{\boldsymbol{Y}}\overset{\infty}{\underset{n=2}{\prod}}{\rm e}^{P_{n}(\boldsymbol{X},\boldsymbol{Y})}$
where $P_{n}(\boldsymbol{X},\boldsymbol{Y})$ is a Lie polynomial
$\boldsymbol{X}$and $\boldsymbol{Y}$ of degree $n$) which is a
dual representation of the well known, Baker-Campbell-Hausdorff theorem\cite{bose}
(${\rm e}^{\boldsymbol{X}}{\rm e}^{\boldsymbol{Y}}={\rm e}^{\boldsymbol{Z}}$,
with $\boldsymbol{Z}=\boldsymbol{X}+\boldsymbol{Y}+\overset{\infty}{\underset{m=2}{\sum}}Q_{m}(\boldsymbol{X},\boldsymbol{Y})$
where $Q_{m}(\boldsymbol{X},\boldsymbol{Y})$ is a Lie polynomial
in $\boldsymbol{X}$and $\boldsymbol{Y}$ of degree $m$).

Now let us define the following system operator $\mathbb{W}={\rm e}^{-\beta\boldsymbol{\mathcal{H}}}$.
Using the Zassenhaus formula\cite{casas} with $N$ operators to obtain
the correction of decoration transformation, as described in Appendix
C, up to second order term. Thus, the equivalent reduced operator
$\mathbb{W}_{r}$, after some algebraic manipulation can be expressed
by the following relation, 
\begin{alignat}{1}
\mathbb{W}_{r}= & {\rm e}^{-\beta(\overset{N}{\underset{i=1}{\sum}}\tilde{\boldsymbol{H}}_{2i-1,2i+1})}+\tfrac{\beta^{2}}{2}\sum_{j=1}^{N-1}\left([\tilde{\boldsymbol{H}}_{2j-1,2j+1},\tilde{\boldsymbol{H}}_{2j+1,2j+3}]-[\boldsymbol{H}'_{2j-1,2j+1},\boldsymbol{H}''_{2j+1,2j+3}]\right)+\mathcal{O}(\beta^{3}),\label{eq:W-fin}
\end{alignat}
 where the second term of eq. \eqref{eq:W-fin} corresponds to the
correction in order $\beta^{2}$. 

For most of the Heisenberg model, the second order term is identically
null, because it involves the only bilinear like power operators.
However, for non-bilinear operators the second order coefficient of
$\beta^{2}$ could be relevant. 

At first glance, one could believe to include this term to correct
our results of decoration transformation. But, we face a serious problem,
because this term includes three nearest neighbor couplings. Consequently,
the spirit of decoration transformation fails, and we cannot to map
into another effective Heisenberg model with only nearest neighbor
coupling term (a simple structure). However, one can use the correction
term just to quantify the validity of our result.

Similarly, for standard Heisenberg model the third order correction
($\beta^{3}$) will be null because the Heisenberg spins are traceless.
Then, this term also does not contribute, unless for non-bilinear
like Hamiltonians, this term could be relevant. One can find an expression
similar to eq. \eqref{eq:W-fin}, but the result will be useless for
decoration transformation, because it involves next nearest neighbor
and further nearest neighbor couplings. 

Another interesting way to prove this correction can be also find
using the cumulant expansion \cite{dunn}, or even alternatively following
the series expansion developed in reference \cite{teresa}, particularly
the last one could be useful for higher order terms.

Note that, for higher dimension the spin models can be mapped in a
similar way, although the coupling terms could be a bit more cumbersome
task.

\subsection{Heisenberg chain as a decorated Heisenberg chain}

As a first application let as consider the spin-1/2 Heisenberg chain,
to verify the quantum decoration transformation approach. Let us consider
a spin-$\frac{1}{2}$ Heisenberg chain with $N'=2N$ sites, and periodic
boundary condition, given by the following Hamiltonian
\begin{equation}
\boldsymbol{\mathcal{H}}=-\sum_{i=1}^{N}\left\{ J\left(\sigma_{2i-1}^{x}s_{2i}^{x}+\sigma_{2i-1}^{y}s_{2i}^{y}+s_{2i}^{x}\sigma_{2i+1}^{x}+s_{2i}^{y}\sigma_{2i+1}^{y}\right)+\Delta\sigma_{2i-1}^{z}s_{2i}^{z}+\Delta s_{2i}^{z}\sigma_{2i+1}^{z}\right\} ,\label{eq:Z-org}
\end{equation}
where we label only for convenience the even site as $\boldsymbol{s}_{2i}=\{s_{2i}^{x},s_{2i}^{y},s_{2i}^{z}\}$
and the odd site by $\boldsymbol{\sigma}_{2i+1}=\{\sigma_{2i+1}^{x},\sigma_{2i+1}^{y},\sigma_{2i+1}^{z}\}$,
but both of them are spin-$\frac{1}{2}$ particles. Thus, we will
call the spin $\boldsymbol{s}_{2i}$ as \textquotedbl{}decorated\textquotedbl{}
spin. Performing the decoration transformation presented in the previous
section, the effective Hamiltonian becomes 
\begin{alignat}{2}
\tilde{\boldsymbol{\mathcal{H}}}= & -N\tilde{J}_{0}-\sum_{j=1}^{N}\left\{ \tilde{J}\left(\sigma_{j}^{x}\sigma_{j+1}^{x}+\sigma_{j}^{y}\sigma_{j+1}^{y}\right)+\tilde{\Delta}\sigma_{j}^{z}\sigma_{j+1}^{z}\right\} , & = & -N\tilde{J}_{0}+\tilde{\boldsymbol{\mathcal{H}}}_{ex}.\label{eq:Z-eff}
\end{alignat}
where $\tilde{\boldsymbol{\mathcal{H}}_{ex}}$ corresponds to standard
Heisenberg model with effective parameters $\tilde{J}_{0}$, $\tilde{J}$
and $\tilde{\Delta}$ which are given by eq. \eqref{eq:tilde-Js}.

This model can be solved numerically for finite size chain, through
exact numerical diagonalization. We choose this approach in order
to confront the exact numerical results and using the quantum decoration
transformation approach. 

Therefore, we can find the free energy per site ($N'=2N$) as 
\begin{equation}
f^{(2N)}=-\frac{1}{2N\beta}\ln\left(Z_{2N}\right),\quad\text{and}\quad\tilde{f}^{(N)}=-\frac{\tilde{J}_{0}}{2}-\frac{1}{2N\beta}\ln\left(\tilde{Z}_{ex,N}\right),
\end{equation}
where $Z_{2N}={\rm tr}\left({\rm e}^{-\beta\boldsymbol{\mathcal{H}}}\right)$
and $\tilde{Z}_{ex,N}={\rm tr}\left({\rm e}^{-\beta\tilde{\boldsymbol{\mathcal{H}}}_{ex}}\right)$,
are the partition functions of Hamiltonian \eqref{eq:Z-org} and \eqref{eq:Z-eff}
respectively.

\begin{table}
\centering{}%
\begin{tabular}{|c|c|c|c|c|c|c|c|}
\hline 
$J$ & $T$ & ($N=4$) $\tilde{f}^{(4)}$  & ($N'=8$) $f^{(8)}$  & $f^{(8)}-\tilde{f}^{(4)}$ & ($N=8$) $\tilde{f}^{(8)}$  & ($N'=16$) $f^{(16)}$  & $f^{(16)}-\tilde{f}^{(8)}$\tabularnewline
\hline 
\hline 
\multirow{5}{*}{$0.1$} & $0.1$ & $-0.2586958$ & $-0.2611776$ & $2.4819\times10^{-3}$ & $-0.2543836$ & $-0.2568705$ & $2.4870\times10^{-3}$\tabularnewline
\cline{2-8} 
 & $0.2$ & $-0.2718840$ & $-0.2742885$ & $2.4045\times10^{-3}$ & $-0.2669289$ & $-0.2693436$ & $2.4147\times10^{-3}$\tabularnewline
\cline{2-8} 
 & $0.3$ & $-0.3040827$ & $-0.3059479$ & $1.8652\times10^{-3}$ & $-0.3024891$ & $-0.3043023$ & $1.8133\times10^{-3}$\tabularnewline
\cline{2-8} 
 & $0.4$ & $-0.3518063$ & $-0.3531251$ & $1.3189\times10^{-3}$ & $-0.3513949$ & $-0.3526862$ & $1.2913\times10^{-3}$\tabularnewline
\cline{2-8} 
 & $0.5$ & $-0.4073722$ & $-0.4083200$ & $0.9479\times10^{-3}$ & $-0.4072546$ & $-0.4081909$ & $0.9364\times10^{-3}$\tabularnewline
\hline 
\hline 
\multirow{5}{*}{$0.5$} & $0.1$ & $-0.2651621$ & $-0.3181141$ & $5.2952\times10^{-2}$ & $-0.2604893$ & $-0.3134708$ & $5.29815\times10^{-2}$\tabularnewline
\cline{2-8} 
 & $0.2$ & $-0.2820743$ & $-0.3304106$ & $4.83363\times10^{-2}$ & $-0.2793074$ & $-0.3272538$ & $4.79464\times10^{-2}$\tabularnewline
\cline{2-8} 
 & $0.3$ & $-0.318616$ & $-0.3565787$ & $3.79625\times10^{-2}$ & $-0.3180631$ & $-0.3554160$ & $4.79464\times10^{-2}$\tabularnewline
\cline{2-8} 
 & $0.4$ & $-0.3674966$ & $-0.3957681$ & $2.82715\times10^{-2}$ & $-0.3673978$ & $-0.3954009$ & $2.80031\times10^{-2}$\tabularnewline
\cline{2-8} 
 & $0.5$ & $-0.4227335$ & $-0.4439272$ & $2.11937\times10^{-2}$ & $-0.4227128$ & $-0.4438067$ & $2.10939\times10^{-2}$\tabularnewline
\hline 
\hline 
\multirow{5}{*}{$1.0$} & $0.1$ & $-0.3200886$ & $-0.4565879$ & $1.36499\times10^{-1}$ & $-0.311581$ & $-0.4476710$ & $1.3609\times10^{-1}$\tabularnewline
\cline{2-8} 
 & $0.2$ & $-0.3381643$ & $-0.4619391$ & $1.23775\times10^{-1}$ & $-0.334471$ & $-0.4571226$ & $1.2265\times10^{-1}$\tabularnewline
\cline{2-8} 
 & $0.3$ & $-0.3727897$ & $-0.4775811$ & $1.04791\times10^{-1}$ & $-0.371347$ & $-0.4755115$ & $1.0416\times10^{-1}$\tabularnewline
\cline{2-8} 
 & $0.4$ & $-0.4182362$ & $-0.5036855$ & $8.54493\times10^{-2}$ & $-0.4175997$ & $-0.5028689$ & $8.5269\times10^{-2}$\tabularnewline
\cline{2-8} 
 & $0.5$ & $-0.4707678$ & $-0.5388776$ & $6.81098\times10^{-2}$ & $-0.4704224$ & $-0.5385579$ & $6.8136\times10^{-2}$\tabularnewline
\hline 
\end{tabular}\caption{\label{tab:1}Free energy as a function of temperature for fixed parameter
$\Delta=1.0$.}
\end{table}

In table \ref{tab:1}, we show the numerical results for fixed parameters
$J=0.1$ and $\Delta=1.0$ (a quasi-Ising model), comparing for a
range of temperatures given in the first column, the second column
presents the free energy per site $\tilde{f}^{(4)}$ for effective
Heisenberg model with $N=4$, the third column shows the free energy
$f^{(8)}$ numerical result for ($N'=2N=8$), and in the fourth column
is shown the difference between both free energies. Whereas the fifth
column displays the free energy per site $\tilde{f}^{(8)}$ of effective
lattice ($N=8$), and in the sixth column, we show the original lattice
free energy $f^{(16)}$ numerical result ($N'=2N=16$). In table \ref{tab:1},
we observe the results are in agreement, and the effective free energy
is slightly higher than original Heisenberg chain, this discrepancy
was to be expected, because the method is only approximate. Notice
that, the effective Heisenberg model only needs half sites compared
to the original Hamiltonian. We have compared only in relatively low
temperature region, this difference is more significant when the temperature
decreases, whereas in the high temperature region both results are
obviously accurate. Surely, this result could be valuable combining
with numerical approaches. In table \ref{tab:1}, we observe the numerical
results for $J=0.5$ is poorly accurate, as well as for $J=1.0$ our
results are even worse, because the quantum coupling is rather relevant.

\textcolor{black}{The above process resembles a real-space renormalization-group
transformation\cite{efrati}, because we are considering uniform spin-1/2
Heisenberg model. However, for mixed or real decorated Heisenberg
model, the decorated transformation goes beyond the renormalization
transformation.}

More detailed analyses using this method would be interesting, but
these analyses are out of the scope of this work.

\subsection{Bond alternating Ising-Heisenberg chain}

As a second application, let us consider the bond alternating Ising-Heisenberg
chain early proposed by Lieb-Schultz-Mattis\cite{Lieb}. Certainly,
this model cannot be mapped exactly into another effective model through
classical decoration transformation. Although the quantum decoration
transformation cannot be applied exactly, here we present an approximate
solution for this model, in the limit of quasi Ising model.

The corresponding Hamiltonian that describes the Ising-Heisenberg
model could be written in a similar way to eq.\eqref{eq:Ham-tot}.
Thus, the Hamiltonian is given by 
\begin{equation}
\boldsymbol{\mathcal{H}}=\sum_{i=1}^{N}\boldsymbol{H}_{2i-1,2i+1},\quad\text{where}\quad\boldsymbol{H}_{2i-1,2i+1}=-\Delta\sigma_{2i-1}^{z}s_{2i}^{z}-J\left(s_{2i}^{x}\sigma_{2i+1}^{x}+s_{2i}^{y}\sigma_{2i+1}^{y}\right)-J_{z}s_{2i}^{z}\sigma_{2i+1}^{z},\label{eq:His-hs-T}
\end{equation}
with periodic boundary condition. The eigenvalues of the Hamiltonian
$\boldsymbol{H}_{2i-1,2i+1}$ in \eqref{eq:His-hs-T} are given by
\begin{alignat}{4}
\varepsilon_{1(2)}= & \frac{-\Delta}{4}-\frac{J_{z}}{4}, & \quad\varepsilon_{3(4)}= & \frac{\Delta}{4}-\frac{J_{z}}{4},\quad & \varepsilon_{5(6)}= & \frac{J_{z}}{4}-\frac{\Delta}{4\cos\phi}, & \quad\varepsilon_{7(8)}= & \frac{J_{z}}{4}+\frac{\Delta}{4\cos\phi},\label{eq:eig-alt-bond}
\end{alignat}
 where $\phi=\arctan\left(\frac{2J}{\Delta}\right)$. Whereas the
corresponding eigenvectors are 
\begin{alignat}{2}
|u_{1(2)}\rangle= & |\pm,\pm,\pm\rangle, & |u_{3(4)}\rangle= & |\mp,\pm,\mp\rangle,\nonumber \\
|u_{5(6)}\rangle= & \frac{\mp1}{\sqrt{2}}\left(\sqrt{1\pm\cos\phi}|+,-,-\rangle+|-,-,+\rangle\right),\quad & |u_{7(8)}\rangle= & \frac{\mp1}{\sqrt{2}}\left(\sqrt{1\mp\cos\phi}|+,-,-\rangle+|-,-,+\rangle\right).
\end{alignat}

Following the recipes in the previous section, we can obtain the reduced
operator $\boldsymbol{W}_{r}$ given by eq.\eqref{eq:R_r}, this reduced
operator becomes just a diagonal matrix in a natural basis, because
$w_{3}=w_{2}$, as well as for null magnetic field we have $w_{4}=w_{1}$.
Thus, $w_{1}$ and $w_{2}$ are obtained using the eq.\eqref{eq:eig-alt-bond},
as follows 
\begin{alignat}{1}
w_{1}= & {\rm e}^{\beta(\frac{Jz+\Delta}{4})}+{\rm e}^{-\beta\tfrac{J_{z}}{4}}\left[\cosh\left(\tfrac{\beta\Delta}{4\cos(\phi)}\right)+\cosh\left(\tfrac{\beta\Delta}{4\cos(\phi)}\right)\cos(\phi)\right],\nonumber \\
w_{2}= & {\rm e}^{\beta(\frac{Jz-\Delta}{4})}+{\rm e}^{-\beta\tfrac{J_{z}}{4}}\left[\cosh\left(\tfrac{\beta\Delta}{4\cos(\phi)}\right)-\cosh\left(\tfrac{\beta\Delta}{4\cos(\phi)}\right)\cos(\phi)\right].
\end{alignat}

The decorated model can be mapped into an effective spin-1/2 Ising
model, given by 
\begin{gather}
\tilde{\boldsymbol{\mathcal{H}}}=-N\tilde{J}_{0}-\tilde{\Delta}\sum_{i=1}^{N}\sigma_{i}\sigma_{i+1}.
\end{gather}

Where the eigenvalues and eigenvectors of $H_{i,i+1}=-\tilde{J_{0}}-\tilde{\Delta}\sigma_{i}\sigma_{i+1}$,
read as 
\begin{alignat}{2}
\tilde{\varepsilon}_{1(2)}=-\tilde{J}_{0}-\frac{\tilde{\Delta}}{4}\longrightarrow & \mid v_{1(2)}\rangle=\mid\pm\pm\rangle,\qquad & \tilde{\varepsilon}_{3(4)}=-\tilde{J}_{0}+\frac{\tilde{\Delta}}{4}\longrightarrow & \mid v_{3(4)}\rangle=\mid\pm\mp\rangle.
\end{alignat}
Obviously, the corresponding $\tilde{\boldsymbol{W}}$ operator 3
is also a diagonal matrix, and each element is compared $w_{i}=\tilde{w}_{i}$.
Thus, the effective parameters are related by 
\begin{gather}
\tilde{J}_{0}=\frac{1}{2\beta}\ln\left(w_{1}w_{2}\right),\qquad\tilde{\Delta}=\frac{2}{\beta}\ln\left(\frac{w_{1}}{w_{2}}\right).
\end{gather}

The spin-1/2 Ising model is a well known model, which can be solved
exactly through the transfer matrix approach. Here we skip the detailed
solution of this model, and only we present the result of the model.
The $2\times2$ transfer matrix can be constructed $\boldsymbol{V}=\left[\begin{array}{cc}
w_{1} & w_{2}\\
w_{2} & w_{1}
\end{array}\right]$, whose eigenvalue are given by $\varLambda_{\pm}={\rm e}^{\beta\tilde{J}_{0}}\left({\rm e}^{\beta\tilde{\Delta}/4}\pm{\rm e}^{-\beta\tilde{\Delta}/4}\right)$.
The partition function of effective Ising chain with spin-1/2, becomes
\begin{gather}
Z_{N}=\Lambda_{+}^{N}+\Lambda_{-}^{N}.
\end{gather}

In the thermodynamic limit $N\rightarrow\infty$, the free energy
is expressed by 
\begin{gather}
f=-\frac{\tilde{J}_{0}}{2}-\frac{1}{2\beta}\ln\left({\rm e}^{\beta\tilde{\Delta}/4}+{\rm e}^{-\beta\tilde{\Delta}/4}\right)=-\frac{1}{2\beta}\ln\left(w_{1}+w_{2}\right).\label{eq:free-e-alt-IH}
\end{gather}

The free energy given by \eqref{eq:free-e-alt-IH} is an approximate
result for Hamiltonian \eqref{eq:His-hs-T}, which is more consistent
only for small $J$. Thus, this result in the limit of $J\rightarrow0$
must be well described by 
\begin{gather}
f=-\frac{1}{2\beta}\ln\left(2{\rm e}^{\frac{\beta J_{z}}{4}}\cosh(\tfrac{\beta\Delta}{4})+2{\rm e}^{-\frac{\beta J_{z}}{4}}\cosh(\tfrac{\beta\Delta}{4}+\tfrac{\beta J^{2}}{2\Delta})\right).
\end{gather}

The thermodynamics solution of bond alternating Ising-Heisenberg chain
is less studied, and this analysis will be discussed elsewhere. 

The ground state energy and spectral analysis of the Hamiltonian given
by \eqref{eq:His-hs-T} has been discussed in reference \cite{Yao,derzhko-strecka}.
This model exhibits a phase transition at zero temperature when $J_{z}=2J$.
Our approach is unable to detect this phase transition at zero temperature,
because our result is accurate in the limit of $J\rightarrow0$, and
in this limit our result exhibits a trivial phase transition at $J_{z}\rightarrow0$.

\section{Conclusion}

Here we present a quantum version of decoration transformation and
show how this transformation could be applied to Heisenberg type models.
The present transformation is an exact mapping when only one decorated
system composes the system. Here we propose an exact quantum decoration
transformation and showing also interesting properties such as the
persistence of symmetry such as (e.g. XXX decorated model$\leftrightarrow$XXX
effective model), and the symmetry breaking during this transformation
(e.g. XXX decorated model$\leftrightarrow$XXZ effective model).

This transformation could be useful to demonstrate the equivalence
between two quantum spin models. In this work, we present some examples,
such spin-1/2 Heisenberg model, and mixed spin-(1,1/2) Heisenberg
model. Unfortunately, the quantum spin decoration transformation cannot
be used to map exactly into another quantum spin lattice model, because
the operators are non-commutative. However, in the \textquotedbl{}classical\textquotedbl{}
limit it could be possible to perform a mapping to establish the equivalence
between two quantum lattice spin models. To study the validity of
this approach for quantum spin lattice model, we use the Zassenhaus
formula and show the correction of quantum decoration transformation,
when it is applied to the lattice spin models. The correction involves
second nearest neighbor, and further nearest neighbor coupling into
a cumbersome task to establish the equivalence between both lattices.
This correction gives us a valuable information about its contribution,
for most of the Heisenberg type models, this one could be irrelevant
at least up to the third order of correction. 

We applied to finite size Heisenberg chain and compared numerically
our result with its exact numerical result, and our result is consistent.
It is worth to mention that the difference of numerical result is
almost independent of the number of sites. Similarly, we also applied
to bond alternating Ising-Heisenberg chain, obtaining an approximate
result. This result is accurate in the limit of weak $xy$ anisotropy
coupling ($J\rightarrow0$) of bond alternating Ising-Heisenberg model.

\section*{Acknowledgment}

This work was supported by Brazilian agency CAPES, FAPEMIG and CNPq.

\appendix

\section{Explicit representation of the Hamiltonians}

In this Appendix we present some Hamiltonians explicitly in its natural
basis:
\begin{enumerate}
\item Spin-$(\frac{1}{2},\frac{1}{2},\frac{1}{2})$ XXZ Hamiltonian $\boldsymbol{H}$
in natural basis $\{|{}_{s,\sigma_{1},\sigma_{2}}\rangle\}$ becomes,
\begin{equation}
\boldsymbol{H}=\left[\begin{array}{cccccccc}
-\frac{\Delta}{2}-\frac{h_{1}}{2}-h & 0 & 0 & 0 & 0 & 0 & 0 & 0\\
0 & -\frac{h_{1}}{2} & 0 & -\frac{J}{2} & 0 & 0 & 0 & 0\\
0 & 0 & -\frac{h_{1}}{2} & -\frac{J}{2} & 0 & 0 & 0 & 0\\
0 & -\frac{J}{2} & -\frac{J}{2} & \frac{\Delta}{2}-\frac{h_{1}}{2}+h & 0 & 0 & 0 & 0\\
0 & 0 & 0 & 0 & \frac{\Delta}{2}+\frac{h_{1}}{2}-h & -\frac{J}{2} & -\frac{J}{2} & 0\\
0 & 0 & 0 & 0 & -\frac{J}{2} & \frac{h_{1}}{2} & 0 & 0\\
0 & 0 & 0 & 0 & -\frac{J}{2} & 0 & \frac{h_{1}}{2} & 0\\
0 & 0 & 0 & 0 & 0 & 0 & 0 & -\frac{\Delta}{2}+\frac{h_{1}}{2}+h
\end{array}\right].\label{eq:Hhhh}
\end{equation}
\item Spin-$(1,\frac{1}{2},\frac{1}{2})$ XXZ Hamiltonian in natural basis
$\{|{}_{s,\sigma_{1},\sigma_{2}}\rangle\}$, becomes a $12\times12$
matrix. The matrix $\boldsymbol{H}$ can be obtained straightforwardly,
as a composition of block matrices

\begin{alignat}{1}
\mathcal{H}_{2}= & \mathcal{H}_{-2}=[-\Delta],\nonumber \\
\mathcal{H}_{1}= & \left[\begin{array}{ccc}
0 & 0 & \frac{-J}{\sqrt{2}}\\
0 & 0 & \frac{-J}{\sqrt{2}}\\
\frac{-J}{\sqrt{2}} & \frac{-J}{\sqrt{2}} & \Delta
\end{array}\right],\;\mathcal{H}_{-1}=\left[\begin{array}{ccc}
\Delta & \frac{-J}{\sqrt{2}} & \frac{-J}{\sqrt{2}}\\
\frac{-J}{\sqrt{2}} & 0 & 0\\
\frac{-J}{\sqrt{2}} & 0 & 0
\end{array}\right],\;\mathcal{H}_{0}=\left[\begin{array}{cccc}
0 & \frac{-J}{\sqrt{2}} & \frac{-J}{\sqrt{2}} & 0\\
\frac{-J}{\sqrt{2}} & 0 & 0 & \frac{-J}{\sqrt{2}}\\
\frac{-J}{\sqrt{2}} & 0 & 0 & \frac{-J}{\sqrt{2}}\\
0 & \frac{-J}{\sqrt{2}} & \frac{-J}{\sqrt{2}} & 0
\end{array}\right].\label{eq:Hihh}
\end{alignat}
Therefore, the Hamiltonian \eqref{eq:Hmt-ihh} can be expressed as
follows 
\begin{equation}
\boldsymbol{H}=\mathcal{H}_{2}\oplus\mathcal{H}_{1}\oplus\mathcal{H}_{0}\oplus\mathcal{H}_{-1}\oplus\mathcal{H}_{-2}.
\end{equation}

The eigenvalues is given in \eqref{eq:E-eig-ihh}, and the eigenvectors
of the Hamiltonian \eqref{eq:Hmt-ihh} are given by

\begin{alignat}{1}
 & |\eta_{1}\rangle=|_{1,+,+}\rangle,\hspace{4.8cm}|\eta_{12}\rangle=|_{-1,-,-}\rangle.\nonumber \\
 & |\eta_{2}\rangle=\frac{1}{2}\left(|_{1,+,-}\rangle+|{}_{1,-,+}\rangle-\sqrt{2}|_{0,+,+}\rangle\right),\qquad|\eta_{9}\rangle=\frac{1}{2}\left(|_{-1,+,-}\rangle+|{}_{-1,-,+}\rangle-\sqrt{2}|_{0,-,-}\rangle\right),\nonumber \\
 & |\eta_{3}\rangle=\frac{1}{2}\left(|_{1,+,-}\rangle+|{}_{1,-,+}\rangle+\sqrt{2}|_{0,+,+}\rangle\right),\qquad|\eta_{10}\rangle=\frac{1}{2}\left(|_{-1,+,-}\rangle+|{}_{-1,-,+}\rangle+\sqrt{2}|_{0,-,-}\rangle\right),\nonumber \\
 & |\eta_{4}\rangle=\frac{1}{\sqrt{2}}\left(|_{1,-,+}\rangle-|{}_{1,+,-}\rangle\right),\hspace{2.5cm}|\eta_{11}\rangle=\frac{1}{\sqrt{2}}\left(|_{-1,-,+}\rangle-|{}_{-1,+,-}\rangle\right),\nonumber \\
 & |\eta_{5}\rangle=\frac{1}{2}\left[\sqrt{1-\cos(\phi)}\left(|_{0,+,-}\rangle+|{}_{0,-,+}\rangle\right)-\sqrt{1+\cos(\phi)}\left(|_{1,-,-}\rangle+|{}_{-1,+,+}\rangle\right)\right],\nonumber \\
 & |\eta_{6}\rangle=\frac{1}{2}\left[\sqrt{1+\cos(\phi)}\left(|_{0,+,-}\rangle+|{}_{0,-,+}\rangle\right)+\sqrt{1-\cos(\phi)}\left(|_{1,-,-}\rangle+|{}_{-1,+,+}\rangle\right)\right],\nonumber \\
 & |\eta_{7}\rangle=\frac{1}{\sqrt{2}}\left(|_{1,-,-}\rangle-|{}_{-1,+,+}\rangle\right),\hspace{2.3cm}|\eta_{8}\rangle=\frac{1}{\sqrt{2}}\left(|_{0,-,+}\rangle-|{}_{0,+,-}\rangle\right),\label{eq:eig-vc2.2}
\end{alignat}

\item Spin-$(\frac{1}{2},\frac{1}{2},\frac{1}{2},\frac{1}{2})$ Hamiltonian\eqref{eq:Ham-3h}
in natural $\{|{}_{s,\sigma_{1},\sigma_{2},\sigma_{3}}\rangle\}$,
can be expressed as $16\times16$ matrix,

The eigenvalues of the Hamiltonian are expressed in \eqref{eq:Eig-E-3h},
and the corresponding eigenvectors are given by
\end{enumerate}
\begin{alignat}{1}
 & |\eta_{1}\rangle=|_{\downarrow,-,-,-}\rangle,\qquad|\eta_{2}\rangle=|_{\uparrow,+,+,+}\rangle,\nonumber \\
 & |\eta_{3}\rangle=\frac{1}{\sqrt{2}}\left(|_{\uparrow,+,-,+}\rangle-|_{\uparrow,+,+,-}\rangle\right),|\eta_{4}\rangle=\frac{1}{\sqrt{2}}\left(|_{\downarrow,-,-,+}\rangle-|_{\downarrow,+,-,-}\rangle\right),\nonumber \\
 & |\eta_{5}\rangle=\frac{1}{\sqrt{6}}\left(|_{\uparrow,+,+,-}\rangle+|_{\uparrow,+,-,+}\rangle-2|_{\uparrow,-,+,+}\rangle\right),|\eta_{6}\rangle=\frac{1}{\sqrt{6}}\left(|_{\downarrow,+,-,-}\rangle+|_{\downarrow,-,+,-}\rangle-2|_{\downarrow,-,-,+}\rangle\right),\nonumber \\
 & |\eta_{7}\rangle=\frac{1}{\sqrt{6}}\left(-|_{\uparrow,+,-,-}\rangle-|_{\uparrow,-+,-}\rangle-|_{\uparrow,-,-,+}\rangle+|_{\downarrow,+,+,-}\rangle+|_{\downarrow,+-,+}\rangle+|_{\downarrow,-,+,+}\rangle\right),\nonumber \\
 & |\eta_{8}\rangle=\frac{1}{\sqrt{6}}\left(|_{\uparrow,+,-,-}\rangle+|_{\uparrow,-+,-}\rangle+|_{\uparrow,-,-,+}\rangle+|_{\downarrow,+,+,-}\rangle+|_{\downarrow,+-,+}\rangle+|_{\downarrow,-,+,+}\rangle\right),\nonumber \\
 & |\eta_{9}\rangle=\frac{1}{2}\left(|_{\uparrow,+,-,-}\rangle-|_{\uparrow,-,-,+}\rangle-|_{\downarrow,+,+,-}\rangle+|_{\downarrow,-,+,+}\rangle\right),|\eta_{11}\rangle=\frac{1}{2}\left(|_{\uparrow,+,-,-}\rangle-|_{\uparrow,-,-,+}\rangle+|_{\downarrow,+,+,-}\rangle-|_{\downarrow,-,+,+}\rangle\right),\nonumber \\
 & |\eta_{10}\rangle=\frac{1}{2\sqrt{3}}\left(|_{\uparrow,+,-,-}\rangle-2|_{\uparrow,-+,-}\rangle+|_{\uparrow,-,-,+}\rangle+|_{\downarrow,+,+,-}\rangle-2|_{\downarrow,+-,+}\rangle+|_{\downarrow,-,+,+}\rangle\right),\nonumber \\
 & |\eta_{12}\rangle=\frac{1}{2\sqrt{3}}\left(|_{\uparrow,+,-,-}\rangle-2|_{\uparrow,-+,-}\rangle+|_{\uparrow,-,-,+}\rangle-|_{\downarrow,+,+,-}\rangle+2|_{\downarrow,+-,+}\rangle-|_{\downarrow,-,+,+}\rangle\right),\nonumber \\
 & |\eta_{13}\rangle=\frac{1}{\sqrt{6}}\left[\sqrt{3(1+\cos(\phi))}|_{\uparrow,-,-,-}\rangle-\sqrt{1-\cos(\phi)}\left(|_{\downarrow,-,+,-}\rangle+|_{\downarrow,+,-,-}\rangle+|_{\downarrow,-,-,+}\rangle\right)\right],\nonumber \\
 & |\eta_{14}\rangle=\frac{1}{\sqrt{6}}\left[\sqrt{3(1+\cos(\phi))}|_{\downarrow,+,+,+}\rangle-\sqrt{1-\cos(\phi)}\left(|_{\uparrow,+,-,+}\rangle+|_{\uparrow,+,+,-}\rangle+|_{\uparrow,-,+,+}\rangle\right)\right],\nonumber \\
 & |\eta_{15}\rangle=\frac{1}{\sqrt{6}}\left[\sqrt{3(1-\cos(\phi))}|_{\uparrow,-,-,-}\rangle+\sqrt{1+\cos(\phi)}\left(|_{\downarrow,-,+,-}\rangle+|_{\downarrow,+,-,-}\rangle+|_{\downarrow,-,-,+}\rangle\right)\right],\nonumber \\
 & |\eta_{16}\rangle=\frac{1}{\sqrt{6}}\left[\sqrt{3(1-\cos(\phi))}|_{\downarrow,+,+,+}\rangle+\sqrt{1+\cos(\phi)}\left(|_{\uparrow,+,-,+}\rangle+|_{\uparrow,+,+,-}\rangle+|_{\uparrow,-,+,+}\rangle\right)\right].\label{eq:eigvct-3h}
\end{alignat}

\section{Diagonalization of the Hamiltonian XXZ with spin-$(1,\frac{1}{2},\frac{1}{2},\frac{1}{2})$}

Conveniently, the Hamiltonian \eqref{eq:TS-1} can be expressed using
the conservation magnet momenta, resulting in block matrices as follows
\begin{equation}
\boldsymbol{H}=H_{\frac{5}{2}}\oplus H_{\frac{3}{2}}\oplus H_{\frac{1}{2}}\oplus H_{-\frac{1}{2}}\oplus H_{-\frac{3}{2}}\oplus H_{-\frac{5}{2}}.
\end{equation}

Where, each block matrices can be described as follows:
\begin{enumerate}
\item For magnetic moment $m=\frac{5}{2}$, there is only one configuration
$|_{+1,+,+,+}\rangle$ for this magnetic moment, $H_{\frac{5}{2}}=[-\frac{3\Delta}{2}].$
Obviously, the corresponding eigenvalues and eigenvectors,
\begin{equation}
\varepsilon_{1}=-\frac{3\Delta}{2},\quad\rightarrow\quad|v_{1}\rangle=|{}_{+1,+,+,+}\rangle.
\end{equation}
\item For magnetic moment $m=\frac{3}{2}$, we have four states given by
$\{|{}_{+1,+,+,-}\rangle$,$|_{+1,+,-,+}\rangle$, $|_{+1,-,+,+}\rangle$,$|_{0,+,+,+}\rangle\}$,
and in this basis the block Hamiltonian can be expressed as
\begin{alignat}{1}
H_{\frac{3}{2}}= & \left[\begin{array}{cccc}
-\frac{\Delta}{2} & 0 & 0 & -\frac{J}{\sqrt{2}}\\
0 & -\frac{\Delta}{2} & 0 & -\frac{J}{\sqrt{2}}\\
0 & 0 & -\frac{\Delta}{2} & -\frac{J}{\sqrt{2}}\\
-\frac{J}{\sqrt{2}} & -\frac{J}{\sqrt{2}} & -\frac{J}{\sqrt{2}} & 0
\end{array}\right].
\end{alignat}
For the isotropic case $J=\Delta$, the eigenvalues and eigenvectors
are given by
\begin{alignat*}{3}
 & \varepsilon_{2}=-\frac{3\Delta}{2}, &  & \rightarrow\quad &  & |v_{2}\rangle=\frac{1}{\sqrt{5}}\left(|_{+1,+,+,-}\rangle+|{}_{+1,+,-,+}\rangle+|{}_{+1,-,+,+}\rangle+\sqrt{2}|{}_{0,+,+,+}\rangle\right),\\
 & \varepsilon_{3}=\Delta, &  & \rightarrow\quad &  & |v_{3}\rangle=\frac{1}{\sqrt{15}}\left(\sqrt{2}|{}_{+1,+,+,-}\rangle+\sqrt{2}|{}_{+1,+,-,+}\rangle+\sqrt{2}|{}_{+1,-,+,+}\rangle-3|{}_{0,+,+,+}\rangle\right),\\
 & \varepsilon_{4}=-\frac{\Delta}{2} &  & \rightarrow\quad &  & |v_{4}\rangle=\frac{1}{\sqrt{2}}\left(|_{+1,+,+,-}\rangle-|{}_{+1,+,-,+}\rangle\right),\\
 & \varepsilon_{5}=-\frac{\Delta}{2} &  & \rightarrow\quad &  & |v_{5}\rangle=\frac{1}{\sqrt{2}}\left(|_{+1,+,+,-}\rangle-|{}_{+1,-+,+}\rangle\right).
\end{alignat*}
\item For magnetic moment $m=\frac{1}{2}$, we have 7 states given by \{$|_{+1,+,-,-}\rangle$,$|_{+1,-,+,-}\rangle$,$|_{+1,-,-,+}\rangle$,$|_{0,+,+,-}\rangle$,$|_{0,+,-,+}\rangle$,
$|_{0,-,+,+}\rangle$,$|_{-1,+,+,+}\rangle$\}, then the block Hamiltonian
becomes
\begin{alignat}{1}
H_{\frac{1}{2}}= & \left[\begin{array}{ccccccc}
\frac{\Delta}{2} & 0 & 0 & -\frac{J}{\sqrt{2}} & -\frac{J}{\sqrt{2}} & 0 & 0\\
0 & \frac{\Delta}{2} & 0 & \frac{J}{\sqrt{2}} & 0 & -\frac{J}{\sqrt{2}} & 0\\
0 & 0 & \frac{\Delta}{2} & 0 & \frac{J}{\sqrt{2}} & -\frac{J}{\sqrt{2}} & 0\\
-\frac{J}{\sqrt{2}} & -\frac{J}{\sqrt{2}} & 0 & 0 & 0 & 0 & -\frac{J}{\sqrt{2}}\\
-\frac{J}{\sqrt{2}} & 0 & -\frac{J}{\sqrt{2}} & 0 & 0 & 0 & -\frac{J}{\sqrt{2}}\\
0 & -\frac{J}{\sqrt{2}} & -\frac{J}{\sqrt{2}} & 0 & 0 & 0 & -\frac{J}{\sqrt{2}}\\
0 & 0 & 0 & -\frac{J}{\sqrt{2}} & -\frac{J}{\sqrt{2}} & -\frac{J}{\sqrt{2}} & \frac{3\Delta}{2}
\end{array}\right].
\end{alignat}
Assuming $J=\Delta$, the corresponding eigenvalues and eigenvectors
read as
\begin{alignat}{3}
 & \varepsilon_{6}=-\frac{3\Delta}{2}, &  & \rightarrow & |v_{6}\rangle= & \frac{1}{\sqrt{10}}\left\{ |_{+1,+,-,-}\rangle+|{}_{+1,-,+,-}\rangle+|{}_{+1,-,-,+}\rangle+\right.\nonumber \\
 &  &  &  &  & \left.+\sqrt{2}(|{}_{0,+,+,-}\rangle+|{}_{0,+,-,+}\rangle+|{}_{0,-,+,+}\rangle)+|{}_{-1,+,+,+}\rangle\right\} ,\nonumber \\
 & \varepsilon_{7}=-\frac{\Delta}{2}, &  & \rightarrow & |v_{7}\rangle= & \frac{1}{\sqrt{6}}\left(|_{+1,+,-,-}\rangle-|{}_{+1,-,-,+}\rangle+\sqrt{2}(|{}_{0,+,+,-}\rangle-|{}_{0,-,+,+}\rangle)\right),\nonumber \\
 & \varepsilon_{8}=-\frac{\Delta}{2}, &  & \rightarrow & |v_{8}\rangle= & \frac{1}{3\sqrt{2}}\left\{ |_{+1,+,-,-}\rangle-2|{}_{+1,-,+,-}\rangle+|{}_{+1,-,-,+}\rangle+\right.\nonumber \\
 &  &  &  &  & \left.+\sqrt{2}(-|{}_{0,+,+,-}\rangle+2|{}_{0,+,-,+}\rangle-|{}_{0,-,+,+}\rangle)\right\} ,\nonumber \\
 & \varepsilon_{9}=\Delta, &  & \rightarrow & |v_{9}\rangle= & \frac{1}{\sqrt{6}}\left(\sqrt{2}(|{}_{+1,+,-,-}\rangle-|{}_{+1,-,-,+}\rangle)-(|{}_{0,+,+,-}\rangle-|{}_{0,-,+,+}\rangle)\right),\nonumber \\
 & \varepsilon_{10}=\Delta, &  & \rightarrow & |v_{10}\rangle= & \frac{1}{3\sqrt{2}}\left\{ \sqrt{2}(|{}_{+1,+,-,-}\rangle-2|{}_{+1,-,+,-}\rangle+|{}_{+1,-,-,+}\rangle)+|{}_{0,+,+,-}\rangle-2|{}_{0,+,-,+}\rangle+|{}_{0,-,+,+}\rangle\right\} ,\nonumber \\
 & \varepsilon_{11}=\Delta, &  & \rightarrow & |v_{11}\rangle= & \frac{1}{3\sqrt{10}}\left\{ -4(|_{+1,+,-,-}\rangle+|{}_{+1,-,+,-}\rangle+|{}_{+1,-,-,+}\rangle)+\right.\nonumber \\
 &  &  &  &  & \left.+\sqrt{2}(|{}_{0,+,+,-}\rangle+|{}_{0,+,-,+}\rangle+|{}_{0,-,+,+}\rangle)+6|{}_{-1,+,+,+}\rangle\right\} ,\nonumber \\
 & \varepsilon_{12}=\frac{5\Delta}{2}, &  & \rightarrow & |v_{12}\rangle= & \frac{1}{3\sqrt{2}}\left\{ |_{+1,+,-,-}\rangle+|{}_{+1,-,+,-}\rangle+|{}_{+1,-,-,+}\rangle+\right.\nonumber \\
 &  &  &  &  & \left.-\sqrt{2}(|{}_{0,+,+,-}\rangle+|{}_{0,+,-,+}\rangle+|{}_{0,-,+,+}\rangle)+3|{}_{-1,+,+,+}\rangle\right\} .
\end{alignat}
\item For $m=-\frac{1}{2}$, we also have 7 states, and the Hamiltonian
$H_{-\frac{1}{2}}$ is identical to $H_{\frac{1}{2}}$, but in different
space spanned by \{$|_{-1,+,+,-}\rangle$,$|_{-1,+,-,+}\rangle$,$|_{-1,-,+,+}\rangle$,$|_{0,-,-,+}\rangle$,$|_{0,-,+,-}\rangle$,
$|_{0,+,+,-}\rangle$,$|_{+1,-,-,-}\rangle$\}. Therefore, the eigenvalues
are identical to that given by $\varepsilon_{6}\cdots\varepsilon_{12}$,
and the corresponding eigenvectors also have same structure, using
the symmetry $+\rightarrow-$ and $-\rightarrow+$, we can generate
the corresponding eigenvectors.
\item For $m=-\frac{3}{2}$, we have an equivalent matrix structure to that
$m=\frac{3}{2}$. Thus, the eigenvalues are identical to the case
2. Using the symmetry $+\rightarrow-$ and $-\rightarrow+$, we can
construct the corresponding eigenvectors.
\item Whereas, for $m=-\frac{5}{2}$, we have an equivalent matrix structure
to $m=\frac{5}{2}$, $H_{-\frac{5}{2}}=[-\frac{3\Delta}{2}]$, so
the corresponding eigenvalues and eigenvectors are
\begin{equation}
\varepsilon_{1}=-\frac{3\Delta}{2},\quad\rightarrow\quad|v_{24}\rangle=|{}_{-1,-,-,-}\rangle.
\end{equation}
\end{enumerate}

\section{Quantum decoration transformation correction}

Using the Zassenhaus formula\cite{casas} with $N$ operators, we
can apply up to second order term, which reads as follows
\begin{alignat}{1}
\mathbb{W}={\rm e}^{-\beta(\overset{N}{\underset{i=1}{\sum}}\boldsymbol{H}_{2i-1,2i+1})}= & {\rm e}^{-\beta\boldsymbol{H}_{1,3}}{\rm e}^{-\beta(\boldsymbol{H}_{3,5}+\boldsymbol{H}_{5,7}+\cdots)}{\rm e}^{-\beta^{2}[\boldsymbol{H}_{1,3},\boldsymbol{H}_{3,5}]}\ldots,\nonumber \\
= & {\rm e}^{-\beta\boldsymbol{H}_{1,3}}{\rm e}^{-\beta\boldsymbol{H}_{3,5}}{\rm e}^{-\beta(\boldsymbol{H}_{5,7}+\boldsymbol{H}_{7,9}+\cdots)}{\rm e}^{-\beta^{2}[\boldsymbol{H}_{3,5},\boldsymbol{H}_{5,7}]}{\rm e}^{-\beta^{2}[\boldsymbol{H}_{1,3},\boldsymbol{H}_{3,5}]}\ldots,\\
 & \vdots\nonumber 
\end{alignat}
After applying $N$ times, we have the following expression
\begin{alignat}{1}
\mathbb{W}= & \prod_{i=1}^{N}{\rm e}^{-\beta\boldsymbol{H}_{2i-1,2i+1}}\prod_{j=N-1}^{1}{\rm e}^{-\beta^{2}[\boldsymbol{H}_{2j-1,2j+1},\boldsymbol{H}_{2j+1,2j+3}]}\ldots.\label{eq:W-rel}
\end{alignat}
Therefore, the operator $\mathbb{W}$ is valid at least up to $\mathcal{O}(\beta^{2})$
, then, simplifying eq.\eqref{eq:W-rel} we have
\begin{equation}
\mathbb{W=}\biggl(\prod_{i=1}^{N}{\rm e}^{-\beta\boldsymbol{H}_{2i-1,2i+1}}\biggr)\biggl(1-\tfrac{\beta^{2}}{2}\sum_{j=1}^{N-1}[\boldsymbol{H}_{2j-1,2j+1},\boldsymbol{H}_{2j+1,2j+3}]\biggr)+\mathcal{O}(\beta^{3}),
\end{equation}
alternatively, even one can write as
\begin{equation}
\mathbb{W=}\prod_{i=1}^{N}{\rm e}^{-\beta\boldsymbol{H}_{2i-1,2i+1}}-\tfrac{\beta^{2}}{2}\sum_{j=1}^{N-1}[\boldsymbol{H}_{2j-1,2j+1},\boldsymbol{H}_{2j+1,2j+3}]+\mathcal{O}(\beta^{3}).
\end{equation}

Performing the partial trace over decorated spins (even sites), we
have
\begin{alignat}{1}
\mathbb{W}_{r}= & {\rm tr}_{e}\biggl({\rm e}^{-\beta(\overset{N}{\underset{i=1}{\sum}}\boldsymbol{H}_{2i-1,2i+1})}\biggr)\nonumber \\
= & {\rm tr}_{e}\biggl(\prod_{i=1}^{N}{\rm e}^{-\beta\boldsymbol{H}_{2i-1,2i+1}}\biggr)-\tfrac{\beta^{2}}{2}{\rm tr}_{e}\biggl(\sum_{j=1}^{N-1}[\boldsymbol{H}_{2j-1,2j+1},\boldsymbol{H}_{2j+1,2j+3}]\biggr)+\mathcal{O}(\beta^{3}),
\end{alignat}
 with ${\rm tr}_{e}$ we denote the partial trace over all even sites.
After performing the partial trace over decorated spins, each element
of the product contains just one even site. Thus, the partial trace
over the even site can be distributed through the product,
\begin{alignat}{1}
\mathbb{W}_{r}= & \prod_{i=1}^{N}{\rm tr}_{2i}\left({\rm e}^{-\beta\boldsymbol{H}_{2i-1,2i+1}}\right)-\tfrac{\beta^{2}}{2}\sum_{j=1}^{N-1}{\rm tr}_{e}\left([\boldsymbol{H}_{2j-1,2j+1},\boldsymbol{H}_{2j+1,2j+3}]\right)+\mathcal{O}(\beta^{3}).\label{eq:W-rdf}
\end{alignat}

Furthermore, using the quantum decoration transformation given by
eq.\eqref{eq:Wr} and substituting the partial trace in eq. \eqref{eq:W-rdf},
we have
\begin{alignat}{1}
\mathbb{W}_{r}= & \prod_{i=1}^{N}{\rm e}^{-\beta\tilde{\boldsymbol{H}}_{2i-1,2i+1}}-\tfrac{\beta^{2}}{2}\sum_{j=1}^{N-1}\left([\boldsymbol{H}'_{2j-1,2j+1},\boldsymbol{H}''_{2j+1,2j+3}]\right)+\mathcal{O}(\beta^{3}),\label{eq:W-rdt}
\end{alignat}
where we denote $\boldsymbol{H}'_{2j-1,2j+1}={\rm tr}_{2j}\left(\boldsymbol{H}_{2j-1,2j+1}\right)$
and $\boldsymbol{H}''_{2j+1,2j+3}={\rm tr}_{2j+2}\left(\boldsymbol{H}_{2j+1,2j+3}\right)$
.

The relation below can be obtained in a similar way as obtained for
\eqref{eq:W-rel}. Hence, we have, 
\begin{alignat}{1}
\prod_{i=1}^{N}{\rm e}^{-\beta\tilde{\boldsymbol{H}}_{2i-1,2i+1}}= & {\rm e}^{-\beta(\overset{N}{\underset{i=1}{\sum}}\tilde{\boldsymbol{H}}_{2i-1,2i+1})}\prod_{j=1}^{N-1}{\rm e}^{-\beta^{2}[\tilde{\boldsymbol{H}}_{2j-1,2j+1},\tilde{\boldsymbol{H}}_{2j+1,2j+3}]},\nonumber \\
= & {\rm e}^{-\beta(\overset{N}{\underset{i=1}{\sum}}\tilde{\boldsymbol{H}}_{2i-1,2i+1})}+\tfrac{\beta^{2}}{2}\sum_{j=1}^{N-1}\left([\tilde{\boldsymbol{H}}_{2j-1,2j+1},\tilde{\boldsymbol{H}}_{2j+1,2j+3}]\right)+\mathcal{O}(\beta^{3}).\label{eq:W-invr}
\end{alignat}

Finally, substituting the relation \eqref{eq:W-invr} into \eqref{eq:W-rdf},
the reduced operator $\mathbb{W}_{r}$ is expressed by eq.\eqref{eq:W-fin}.


\begin{thebibliography}{10}
\bibitem{xy}A detailed discussion will be considered later in another
journal.
\end{thebibliography}
\end{document}